\documentclass[12pt]{article}
\usepackage{epsfig}

\makeatletter 
\@addtoreset{equation}{section} 
\makeatother 
\renewcommand{\theequation}{\thesection.\arabic{equation}} 

\def\draftlabel#1{{\@bsphack\if@filesw {\let\thepage\relax
   \xdef\@gtempa{\write\@auxout{\string
      \newlabel{#1}{{\@currentlabel}{\thepage}}}}}\@gtempa
   \if@nobreak \ifvmode\nobreak\fi\fi\fi\@esphack}
        \gdef\@eqnlabel{#1}}
\def\@eqnlabel{}
\def\@vacuum{}
\def\draftmarginnote#1{\marginpar{\raggedright\scriptsize\tt#1}}

\def\numberbysection{\@addtoreset{equation}{section}
        \def\theequation{\thesection.\arabic{equation}}}

\def\titlepage{\@restonecolfalse\if@twocolumn\@restonecoltrue\onecolumn
     \else \newpage \fi \thispagestyle{empty}\c@page\z@
        \def\thefootnote{\fnsymbol{footnote}} }

\def\endtitlepage{\if@restonecol\twocolumn \else \newpage \fi
        \def\thefootnote{\arabic{footnote}}
        \setcounter{footnote}{0}}  

\catcode`@=12
\relax

\def\ap#1#2#3{ {\sl Ann. Phys.\/} {\bf#1}, #2(#3)}
\def\figcap{\section*{Figure Captions\markboth
        {FIGURECAPTIONS}{FIGURECAPTIONS}}\list
        {Figure \arabic{enumi}:\hfill}{\settowidth\labelwidth{Figure
999:}
        \leftmargin\labelwidth
        \advance\leftmargin\labelsep\usecounter{enumi}}}
 \relax

\def\draftlabel#1{{\@bsphack\if@filesw {\let\thepage\relax
   \xdef\@gtempa{\write\@auxout{\string
      \newlabel{#1}{{\@currentlabel}{\thepage}}}}}\@gtempa
   \if@nobreak \ifvmode\nobreak\fi\fi\fi\@esphack}
        \gdef\@eqnlabel{#1}}
\def\@eqnlabel{}
\def\@vacuum{}
\def\draftmarginnote#1{\marginpar{\raggedright\scriptsize\tt#1}}

\def\draft{\oddsidemargin -.5truein
        \def\@oddfoot{\sl preliminary draft \hfil
        \rm\thepage\hfil\sl\today\quad\militarytime}
        \let\@evenfoot\@oddfoot \overfullrule 3pt
        \let\label=\draftlabel
        \let\marginnote=\draftmarginnote
   \def\@eqnnum{(\theequation)\rlap{\kern\marginparsep\tt\@eqnlabel}%
\global\let\@eqnlabel\@vacuum}  }

\def\be{\begin{equation}}
\def\ee{\end{equation}}
\def\ba{\begin{eqnarray}}
\def\ea{\end{eqnarray}}

\newcommand{\beq}{\begin{equation}}
\newcommand{\eeq}[1]{\label{#1}\end{equation}}
\newcommand{\ber}{\begin{eqnarray}}
\newcommand{\eer}[1]{\label{#1}\end{eqnarray}}
\newcommand{\eqn}[1]{(\ref{#1})}

\def\tr{\,{\rm tr}\,}
\def\str{\,{\rm str}\,}
\def\ap{\alpha'}
\def\at{{\tilde \alpha}'}
\def\a{\alpha}
\def\b{\beta}
\def\g{\gamma}
\def\G{\Gamma}
\def\d{\partial}
\def\e{\epsilon}
\def\p{\psi} 
\def\c{\chi}
\def\cb{{\overline\chi}}

\def\m{\mu}
\def\n{\nu}
\def\r{\rho}
\def\l{\lambda}
\def\lb{{\overline\lambda}}
\def\k{\kappa}
\def\s{\sigma}

\def\vf{\varphi}
\def\O{{\cal O}}
\def\ub{{\overline u}}
\def\ks{{k \kern-.5em /}}
\def\es{{\e \kern-.4em /}}
\def\ds{{\partial \kern-.5em /}}
\def\Ds{{D \kern-.6em /}}
\def\gt{{\tilde g}}
\def\ct{{\tilde c}}
\def\gh{{\hat g}}
\def\ch{{\hat c}}
\def\hh{{\hat h}}

\def\no{\noindent}

\def \ha {{1\over 2}}

\def\inv{^{\raise.15ex\hbox{${\scriptscriptstyle -}$}\kern-.05em 1}}

\begin{document}
\newcommand{\fig}[3]{\epsfxsize=#1\epsfysize=#2\epsfbox{#3}}



\pagestyle{empty}
\begin{flushright}
\small
NEIP-01-004\\
{\tt hep-th/0106062}\\
June 2001
\normalsize
\end{flushright}

\begin{center}


\vspace{.7cm}

{\Large {\bf Higher-derivative corrections to the\break

             non-abelian Born-Infeld action}}

\vspace{2.cm}

{\large Adel~Bilal}

\vskip 0.4truecm

Institute of Physics, University of Neuch\^atel\\
   rue Breguet 1, 2000 Neuch\^atel, Switzerland\\
   {\tt adel.bilal@unine.ch}

\vskip 3.0cm


{\bf Abstract}

\end{center}

\begin{quotation}

\small
\noindent
We determine higher-derivative terms in the open superstring effective 
action with ${\rm U}(N)$ gauge group up to and including order $\ap^4$ as can be 
extracted from 4 boson, 2 boson - 2 fermion and 4 fermion string 
scattering amplitudes. This yields corrections to the non-abelian Born-Infeld
action involving higher derivatives as is relevant for studying D-branes 
beyond the slowly varying field approximation. While at order $\ap^2$ the 
action has recently been shown to be a symmetrised trace, 
this no longer is true at order $\ap^3$ or $\ap^4$. 
We argue that these terms including higher derivatives are as important 
in a low-energy expansion as e.g. the much-discussed $\ap^4 F^6$ terms.
In particular a computation of the fluctuation spectra at order $\ap^4$
has to take into account these non-symmetrised trace higher-derivative terms 
computed here.

\end{quotation}

\newpage


\pagestyle{plain}

\setcounter{section}{0}
\section{Introduction\label{Intro}}

The abelian Born-Infeld action \cite{AT} and its supersymmetric generalisation
\cite{Aga} capture the low-energy dynamics of open superstrings without 
Chan-Paton factors or, equivalently, of a single D9-brane. The analogous 
description in the presence of several D-branes is not known. For $N$ 
D$p$-branes the gauge group is ${\rm U}(N)$ and the $9-p$ embedding 
coordinates, on which all background fields depend, are 
${\rm U}(N)$-valued matrices. To avoid this difficulty one can restrict 
oneself to studying D9-branes only. The leading low-energy limit of the 
effective action then is ${\rm U}(N)$ super Yang-Mills theory. There has 
been quite some controverse, however, concerning the higher-order 
corrections. It was proposed that they are again captured by some 
non-abelian generalisation of the Born-Infeld action with the trace  
over the ${\rm U}(N)$-generators being a symmetrised trace \cite{AT2}. 
While this proposal assumed that derivatives of the Yang-Mills field 
strength are small in an appropriate sense and can be neglected, it 
allows for large fields with $\ap F$ of order one. The symmetrised trace 
proposal was challenged at order $\ap^4 F^6$ by comparing fluctuation 
spectra of this non-abelian Born-Infeld action with background magnetic 
fields against a direct calculation of the spectrum from the dual picture 
of D-branes at angles \cite{HT,DST}. There were further indications 
from requiring $\k$-symmetry of the effective action that deviations 
from a symmetrised trace may already occur at order $\ap^2$ for terms 
involving fermions \cite{BRS}.

The most direct way to settle this issue would be to obtain the effective 
action from open superstring scattering amplitudes. While this yielded 
the bosonic terms through order $\ap^2 F^4$ long ago \cite{GW}, the full 
action at this order, including fermions, as deduced from the string 
scattering amplitudes was obtained recently \cite{BBdRS}. At this order, 
the symmetrised trace turns out to be the correct answer, and moreover, 
it is uniquely determined from the scattering amplitudes (up to field 
redefinitions - as always). Independently it was shown \cite{goteborg} 
that imposing linear susy at order $\ap^2$ (almost) uniquely leads to 
the same action. The conflict with $\k$-symmetry was resolved \cite{BBdRS}
by realising that $\k$-symmetry actually fails at orders higher than those
considered in \cite{BRS}.

The expansion of the action can be viewed as an expansion in $\ap$ and 
the Yang-Mills coupling $g$, with $\ap g F_{\m\n}$ being dimensionless. 
Thus the lowest order $F^2$ term can be corrected by $\ap^2 g^2 F^4$, 
$\ap^3 g^3 F^5$, $\ap^4 g^4 F^6$ etc, but also by terms of the form 
$\ap^3 g^2 D^2 F^4$ and $\ap^4 g^2 D^4 F^4$ etc. The standard assumption 
of large but slowly-varying fields is that one can neglect 
$\ap^3 g^2 D^2 F^4$ with respect to $\ap^3 g^3 F^5$ etc. While it was 
always clear that there is some ambiguity here in the non-abelian case 
since $[D,D]\sim g F$, it was argued that the assumption would still be 
valid in a broad class of situations.

I will instead point out that allowing large fields one also has to 
allow large derivatives. There are two simple arguments, one formal 
and one physical. The formal argument is specific to the non-abelian case and
was pointed out to me by Alex Sevrin: 
If one is interested in the equations of motion or in the fluctuation 
spectra rather than in the action itself, one has to vary once or twice 
with respect to $A_\m$. Varying e.g. $\ap^4 g^4 F^6$ twice yields a term 
of the form $\ap^4 g^4 F^2 (D F)^2$. On the other hand, the covariant 
derivatives $D_\m$ also contains $g A_\m$ and varying $\ap^4 g^2 (D F)^4$ 
twice also leads to the same term  $\ap^4 g^4 F^2 (D F)^2$. So there seems 
to be no reason to include only the $F^6$ term and not the $(D F)^4$ term.

The second argument applies to both the non-abelian and the abelian case.
The basic idea is simple: if there is some region ${\cal R}$ in space 
where the fields are large, with $\ap g F$ of order one, they also have 
to fall off to zero far from this region. If they fall off slowly enough 
to have small derivatives, then they actually stay large over a  region 
much bigger than ${\cal R}$. The total energy then is such that this 
configuration forms a black hole, so that gravitational effects can no 
longer be neglected and it is certainly not enough to only consider the 
effective action for the gauge fields. To avoid this scenario, the fields 
would have to fall off quickly enough outside the region ${\cal R}$, 
leading to large gradients with terms like $\ap^4 g^2 (D F)^4$ comparable 
to $\ap^4 g^4 F^6$.

Let me make the latter argument a bit more quantitative. For simplicity, we 
make the argument in the more familiar setting of four dimensions and 
then indicate 
how it extends to arbitrary dimensions. We also drop all irrelevant 
numerical factors of order one. 

The Schwarzschild radius $R_H$ of a mass 
$M$ object is $R_H\sim {M\over M_P^2}$ with $M_p$ being the Planck mass. 
We will assume that the string scale ${1\over \sqrt{\ap}}$ is not very 
different from $M_P$ so that $R_H\sim \ap M$. Consider a Yang-Mills field 
configuration with 
\be\label{fsize}
\ap F\sim \l \ , 
\ee
where $\l$ is a dimensionless parameter 
controlling the strength of the field. Suppose this configuration extends 
over some region and falls to zero within a radius $R$. The derivatives 
then are at least 
\be\label{dersize}
\ap D F \sim  {\l\over R}
\ee 
over part of this region. The total energy contained within this region is 
$M\sim F^2 R^3 \sim {\l^2 R^3\over \ap^2}$ with the Schwarzschild radius 
being $R_H\sim {\l^2 R^3\over \ap}$. This configuration will avoid being 
a black hole if $R>R_H$. i.e. $R\ {<\atop \sim}\ {\sqrt{\ap}\over \l}$. But 
this implies that the derivatives are 
\be\label{derlimit}
\ap^{3/2} D F \ \ {\textstyle{>\atop \sim}}\ \ \l^2
\ee
and cannot be neglected.  Specifically, for various terms
that appear in the action  we get 
\be\label{ordermagnitude}
\ap^4 F^6 \sim {\l^6\over \ap^2} \quad , \quad 
\ap^3 F^2 (D F)^2 \ \ {\textstyle{>\atop \sim}}\ \ {\l^6\over \ap^2}
\quad , \quad 
\ap^4 (D F)^4 \ \ {\textstyle{>\atop \sim}}\ \  {\l^8\over \ap^2} \ .
\ee
Choosing a small $\l$, one can make $\ap^4 (D F)^4$ smaller than 
$\ap^4 F^6$ by a factor $\l^2$. But small $\l$ means small fields, 
somewhat contrary to the assumptions. In any case, the higher-derivative 
term $\ap^3 F^2 (D F)^2$ is exactly as important, if not more, as 
$\ap^4 F^6$, whatever the size of $\l$.

This argument can be extended to any space-time dimension $d$ where 
various powers of $d$ appear in various places, but the qualitative 
conclusion remains the same: To stay within the validity of the whole 
scheme, one should avoid that a black hole is formed. This necessarily 
implies that higher-derivative terms are as important as higher-order 
non-derivative terms.

With this motivation in mind, in this paper, I will determine higher-derivative 
corrections to the non-abelian Born-Infeld action, by working out these 
corrections to the open superstring effective action as it can be obtained 
from four-point scattering amplitudes. The four-point amplitudes are 
well-known and can be easily expanded in $\ap$. It is then not too difficult 
to match these amplitudes against those obtained from an effective action 
including not more than four fields ($F_{\m\n}$ or fermions $\c$, 
so the action 
is of order $g^2$ in the Yang-Mills coupling) but with an arbitrary number 
of derivatives. Thus we determine the complete action at order $\ap^3 g^2$
and at order $\ap^4 g^2$. As it is obvious from the expansion 
of the scattering amplitudes, the effective action at order $\ap^3 g^2$ is 
proportional the the product of two structure constants of the gauge group 
and hence is {\it not} a symmetrised trace. Also, at order $\ap^4 g^2$, there 
are two types of contributions, some again proportional the the product of 
two structure constants, and others being a symmetrised trace.

The plan of this paper is the following: in section 2, we present the 
various four-point string scattering amplitudes, with particular 
emphasis on the various combinations of the traces over the ${\rm U}(N)$ 
generators that appear, and give their expansions up to order $\ap^6$. In 
section 3, we set up the computation of the effective action and briefly 
recall the results of \cite{BBdRS} at order $\ap^2$. In section 4 we compute
the complete action including fermions at order $\ap^3$ (and $\sim g^2$) while 
in section 5 we
obtain the complete action at order $\ap^4$ (and $\sim g^2$).

While the present paper was typed, a paper by Refolli, Santambrogio, Terzi 
and Zanon \cite{RSTZ} appeared which also determines higher-derivative 
correction to the non-abelian Born-Infeld action. While the paper \cite{RSTZ}
only considers 
the bosonic part of the action and only with two derivatives, it also contains 
some information on $F^5$ terms.


\section{Open superstring four-point scattering amplitudes}

In this section we will first review the computation of the tree-level 
open string (disc) four-point amplitudes between the massless gauge 
bosons and their fermionic partners (gauginos) and then study 
in some detail their expansion in powers of $\ap$. There is a 4 boson, 
a 4 fermion and a 2 boson / 2 fermion amplitude. We take the 
external momenta $k_1, \ldots k_4$ all  as incoming, assign 
Chan-Paton labels $a,b,c,d= 1, \ldots {\rm dim\ U}(N)$, and 
wave-functions $u_i$ to the external fermions and polarisations 
$\e_j$ to the external bosons. This is depicted in Fig. 1 for the 
example of a 2 boson / 2 fermion amplitude.

\begin{figure}[h]
\centerline{\epsfig{figure=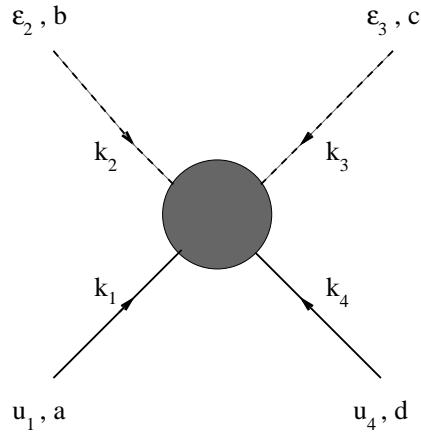, height=6cm, width=6cm}}
\caption{2 boson / 2 fermion scattering amplitude}
\end{figure}

Any of these 4 point amplitudes is a sum of six disc diagrams 
corresponding to the 6 different cyclic orderings of the vertex 
operators as shown in Fig. 2.

\begin{figure}[h]
\centerline{\epsfig{figure=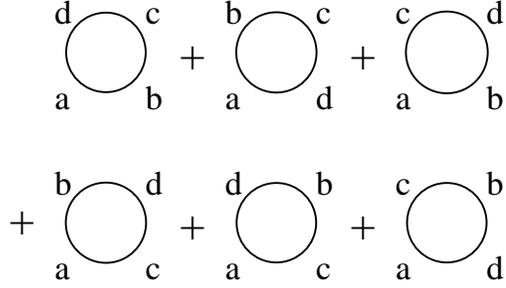, height=4cm, width=4cm}}
\caption{The six different cyclic orderings}
\end{figure}

The contribution of each of the six orderings then is given \cite{GSW,POL} 
by the product of

\no
1.) a trace of the product of matrices $\l_a$ in the fundamental 
representation of ${\rm U}(N)$, taken in the cyclic order given 
by the diagram of Fig. 2, e.g. for the first one: 
$\tr \l_a\l_b\l_c\l_d \equiv t_{abcd}$

\no
2.) a function $G$ depending on the two Mandelstam variables ``flowing" 
through the diagram ``horizontally" and ``vertically". For the first 
diagram of Fig. 2 e.g. the vertical momentum flow gives $(k_1+k_2)^2=s$ 
while the horizontal momentum flow gives $(k_1+k_4)^2=u$. Clearly, 
the 1. and 2. diagram give $G(s,u)$, the 3. and 4. give $G(s,t)$ and 
the 5. and 6. give $G(t,u)$. The function $G$ is given by
\be\label{Gfct}
G(s,t)=\ap^2 {\G(-\ap s) \G(-\ap t)\over \G(1-\ap s-\ap t)}
\ee
and is the same independent of the nature (boson or fermion) of the 
massless external states.

\no
3.) a kinematic factor $K$ depending on the polarisations and 
wave-functions in the given cyclic order as well as on the momenta. 
It is independent of $\ap$. This factor would actually be the same also
for loop amplitudes. In the present example of 2 boson / 2 fermion 
scattering of Fig 1, the 3. diagram of Fig. 2 would e.g. come with a 
$K(u_1, \e_2, u_4, \e_3)$.

\no
4.) a normalisation factor which we will take to be $-8ig^2$.

\no
5.) a minus sign for any diagram in Fig. 2 which differs from the first 
one by the permutation of two fermions. Note that these signs will be 
cancelled in the end by the corresponding antisymmetry of the $K$-factor.

Let us now discuss these ingredients in more detail.

\subsection{The traces}

The traces come in 3 combinations (recall that 
$t_{abcd}\equiv \tr \l_a\l_b\l_c\l_d$):
\be\label{traces}
T_1 = t_{abcd}+t_{dcba} \quad , \quad
T_2 = t_{abdc}+t_{cdba}\quad , \quad
T_3 = t_{acbd}+t_{dbca}\ . 
\ee
Using $[\l_a, \l_b]=i f_{abc} \l_c$ and $\{\l_a, \l_b\}=d_{abc} \l_c$
as well as the normalisation $\tr \l_a \l_b=\delta_{ab}$ it is easy to 
show that
\ba\label{tracecomb}
T_1+T_2+T_3 = 6 \str \l_a\l_b\l_c\l_d 
= \ha\left( d_{abe}d_{cde} + d_{ace}d_{bde} + d_{ade}d_{bce}\right) \cr
T_2-T_1 =f_{abe}f_{cde}\ , \quad
T_2-T_3 =f_{ace}f_{bde}\ , \quad
T_1-T_3 =f_{ade}f_{bce}\ . 
\ea
Using the Jacobi identities of the appendix we also obtain
\ba\label{indivtraces}
T_1&=&{1\over 2} \left(d_{abe}d_{cde}+d_{ade}d_{bce}-d_{ace}d_{bde}\right) \cr
T_2&=&{1\over 2} \left(d_{abe}d_{cde}+d_{ace}d_{bde}-d_{ade}d_{bce}\right) \cr
T_3&=&{1\over 2} \left(d_{ace}d_{bde}+d_{ade}d_{bce}-d_{abe}d_{cde}\right) \ .
\ea

\subsection{The function $G$}

The dependence of the amplitudes on the Mandelstam variables is contained 
in the function $G$ as given in eq. \eqn{Gfct}. Its expansion in powers 
of $\ap$ is,
up to and including $\ap^6$ terms:
\ba\label{Gexpansion}
G(s,t)&=&
{1\over st} -{\pi^2\over 6} \ap^2 +{c_2\over 2} (s+t) \ap^3 
-{\pi^4\over 360} (4s^2+4t^2+st) \ap^4\cr
&+& \left[ {c_4\over 24} (s+t) (s^2+t^2+st) 
- {\pi^2 c_2\over 12} (s+t) st \right] \ap^5\cr
&+&\left[{c_2^2\over 8}(s+t)^2st 
-{\pi^6\over 15120} (16s^4+16t^4+12s^3t+12st^3+23s^2t^2)\right] \ap^6
+\O(\ap^7) \ , 
\ea
where the $c_n$ are defined by
\be\label{polygamma}
c_n={{\rm d}^n \psi(z)\over {\rm d} z^n}\Big\vert_{z=1} = 
{{\rm d}^{n+1} \log\G(z)\over {\rm d} z^{n+1}}\Big\vert_{z=1} \ ,
\ee
and in particular $c_2 = 2 \zeta(3) \simeq -2.40411$ and $c_4\simeq -24.8863$.

Using $s+t=-u$ we may rewrite $G$ in a way which makes explicit the 
terms which are invariant when changing the arguments:
\ba\label{Gexpansion2}
G(s,t)&=&
{1\over st} -{\pi^2\over 6}\,  \ap^2 -\zeta(3) u\,  \ap^3 
-\left[ {\pi^4\over 180} (s^2+t^2+u^2) - {\pi^4\over 120} st\right] \ap^4\cr
&+& \left[ {\pi^2 \zeta(3)\over 6}  stu  
- {c_4\over 48} (s^2+t^2+u^2) u \right] \ap^5\cr
&+&\Bigg[-{\pi^6\over 2880} (s^2+t^2+u^2)^2 
+ {\pi^6\over 3024} (s^2+t^2+u^2) \left( st + {u^2\over 2}\right) \cr
& & \quad
+ \left( {\pi^6\over 3024} + {\zeta(3)^2\over 2}\right)(stu)\, u \Bigg] \ap^6
+\O(\ap^7) \ , 
\ea

\subsection{The kinematical factors $K$}

The $K$ factors are given in ref. \cite{GSW}. Some care has to be exercised
while copying the formula since our conventions are different from those of 
ref. \cite{GSW}\footnote{
The differences are: a) $s_{\rm GSW}=-s$, $t_{\rm GSW}=-u$, $u_{\rm GSW}=-t$,
b) $\{\G^\m,\G^\n\}_{\rm GSW}= - 2 \eta_{\rm GSW}^{\m\n}$ while we take 
$\{\g^\m,\g^\n\}=  2 \eta^{\m\n}$, and c) we also must change the overall  
normalisation by a factor $-{1\over 4}$ for the 4 fermion and the 
2 boson / 2 fermion case, while in the 4 boson case the GSW normalisation 
is appropriate.}.

For 4 boson we get:
\be\label{kbos}
K(\e_1,\e_2,\e_3,\e_4)= - {t u\over 4} \e_1\cdot \e_2\, \e_3 \cdot \e_4
- {s u\over 4} \e_1\cdot \e_3\,  \e_2 \cdot \e_4
- {s t\over 4} \e_1\cdot \e_4\,  \e_2 \cdot \e_3 
-{s\over 2}{\cal K}_s -{t\over 2}{\cal K}_t  -{u\over 2}{\cal K}_u
\ee
where
\ba\label{ks}
{\cal K}_s &=& \e_1\cdot k_4\,  \e_3 \cdot k_2\,  \e_2\cdot \e_4
+ \e_2\cdot k_3\,  \e_4 \cdot k_1\,  \e_1\cdot \e_3 
+ \e_1\cdot k_3\,  \e_4 \cdot k_2\,  \e_2\cdot \e_3
+ \e_2\cdot k_4\,  \e_3 \cdot k_1\,  \e_1\cdot \e_4\cr
{\cal K}_t &=& {\cal K}_s \vert_{2\leftrightarrow 3}\cr
{\cal K}_u &=& {\cal K}_s \vert_{2\leftrightarrow 4}
\ea
Note that $K(\e_1,\e_2,\e_3,\e_4)$ is completely symmetric under any 
permutation $i\leftrightarrow j$ and it vanishes if we replace $\e_i$ 
by $k_i$ as required by gauge invariance.

For four fermions the K-factor is given by
\be\label{kferm}
K(u_1,u_2,u_3,u_4)={s\over 8}\, \ub_1\g_\m u_4\,  \ub_2 \g^\m u_3
- {u\over 8}\,  \ub_1\g_\m u_2\,  \ub_4 \g^\m u_3 \ .
\ee
The $u_i$ are the (commuting) fermion ten-dimensional Majorana-Weyl 
wave-functions. Hence we have 
$\ub_i\g^\m u_j = \ub_j \g^\m u_i$ and the Fierz identity
\be\label{fierz1}
\ub_1\g_\m u_2\,  \ub_3 \g^\m u_4 
+ \ub_1\g_\m u_3\,  \ub_4 \g^\m u_2 
+ \ub_1\g_\m u_4\,  \ub_2 \g^\m u_3  = 0
\ee
which together with the relation $s+t+u=0$ implies that 
$K(u_1,u_2,u_3,u_4)$ is completely antisymmetric under the exchange of any 
two fermions, e.g. we have $ K(u_1,u_2,u_4,u_3)=-K(u_1,u_2,u_3,u_4)$ etc.

For two fermions and two bosons, ref. \cite{GSW} considers two cases separately: 
the two fermions are adjacent or not. Both cases actually lead to the 
{\it same} $K$-factor as we now show. For fermions that are adjacent 
in the cyclic 
order we get from \cite{GSW}
\be\label{kferbos}
K(u_1,\e_2,\e_3,u_4)={u\over 8}  A + {s\over 8} B
\ee
where we define the convenient expressions ($\ks \equiv k_\m \g^\m$)
\ba\label{AB}
A&=& \ub_1 \es_2 (\ks_3+\ks_4)\es_3 u_4 \cr
B&=& 2 \ub_1 \left( \es_3 \, k_3\cdot\e_2 -\es_2 \, k_2\cdot \e_3 
- \ks_3\, \e_2\cdot \e_3 \right) u_4 \ . 
\ea
Using the on-shell properties
\be\label{onshell}
k_2\cdot \e_2 = k_3\cdot\e_3 = \ks_4 u_4 = \ub_1 \ks_1 =0
\ee
one easily shows
\ba\label{ABsym}
A\vert_{2\leftrightarrow 3} &=& A-B \quad , \quad 
A\vert_{1\leftrightarrow 4} = B-A \cr
B\vert_{2\leftrightarrow 3} &=& -B \quad\quad , \quad 
B\vert_{1\leftrightarrow 4} = B \ .
\ea
It then follows that $K(u_1,\e_2,\e_3,u_4)$ is symmetric under the 
exchange of the two bosons and antisymmetric under exchange of the 
two fermions. If the fermions are not adjacent in the cyclic ordering
we get instead from \cite{GSW}
\ba\label{kferbos2}
K(u_1,\e_2,u_4,\e_3)&=&-{t\over 8}\,  \ub_1 \es_2(\ks_3+\ks_4) \es_3 u_4
-{s\over 8}\,  \ub_1 \es_3 (\ks_2+\ks_3)\es_2 u_4 \cr
&=& -{t\over 8} A -{s\over 8} (A-B) ={u\over 8} A +{s\over 8} B
\ea
which is actually identical to the other $K$-factor \eqn{kferbos} 
for adjacent fermions. Thus there is a single $K$-factor for 
2 bosons / 2 fermions, just as for 4 boson or 4 fermion scattering.

These kinematical factors are actually determined by the required 
(anti)symmetry, (linearized) gauge invariance and dimensional reasoning. 
In the 2 fermion / 2 boson case e.g. the (anti)symmetry and dimensional 
reasoning require $K$ to be of the form 
$K_\b={u\over 8} A +{1\over 8} \left( s+ \b \left( t-{u\over 2}\right)\right) B$ and then gauge invariance (vanishing upon $\e_i\to k_i$) fixes $\b=0$.

It follows that any of the four-point (tree-level) amplitudes we are 
interested in takes the form
\ba\label{A4gen}
A_4&=&-8ig^2\ K(1,2,3,4) \ \times\cr
&\times&\left\{
(t_{abcd}+t_{dcba}) G(s,u)
+(t_{abdc}+t_{cdba})G(s,t)
+(t_{acbd}+t_{dbca})G(t,u)  \right\} \ .\cr
& &
\ea
Note that any minus signs introduced when two fermions in Fig. 2 are 
permuted with respect to the reference configuration has been cancelled 
by another minus sign when performing the same permutation on the  
arguments of $K$ to rewrite it as $K(1,2,3,4)$.

\subsection{$\ap$-expansion of the four-point amplitude}

Inserting the $\ap$-expansion of the $G$-function into \eqn{A4gen} we get
for any of the four-point amplitudes 
\be\label{A4exp}
A_4=-8ig^2\ K(1,2,3,4) \sum_{N=0}^\infty a_4^{(n)} \ap^n\ .
\ee
The lowest order term can be written in 3 equivalent ways:
\ba\label{a40}
a_4^{(0)}
&=& {1\over s} \left( {1\over t} f_{ace}f_{bde} 
+ {1\over u} f_{ade}f_{bce}\right) \cr
&=& -{1\over u} \left( {1\over s} f_{abe}f_{cde} 
+ {1\over t} f_{ace}f_{bde}\right) \cr
&=& {1\over t} \left( {1\over s} f_{abe}f_{cde} 
- {1\over u} f_{ade}f_{bce}\right) \ .
\ea
This vanishes in the abelian case: there is no lowest order photon-photon 
scattering. Clearly, there is no order $\ap$ contribution and $a_4^{(1)}=0$.

The obvious fact about the order $\ap^2$ contribution is that it is always 
a symmetrised trace. Indeed, at order $\ap^2$ the function $G$ is just a 
constant, and thus all traces contribute equally, leading to a symmetrised 
trace:
\be\label{a42}
a_4^{(2)}= -\pi^2 \str\l_a\l_b\l_c\l_d \ .
\ee
At order $\ap^3$, since $s+t+u=0$, no symmetrised trace part remains 
and the contribution only contains products of structure constants $f$ 
(as was also the case at order $\ap^0$ for the same reason):
\ba\label{a43}
a_4^{(3)}&=&\zeta(3)\left( t f_{abe}f_{cde} + s f_{ace}f_{bde}\right)\cr
&=&{\zeta(3)\over 3}\left[ (t-u) f_{abe}f_{cde} +(s-u) f_{ace}f_{bde}
+ (s-t) f_{ade}f_{bce} \right]
\ea
which again can be rewritten in various ways.

The order $\ap^4$ contribution is more interesting as it contains a manifestly symmetric and a non-symmetric piece:
\ba\label{a44}
a_4^{(4)}&=& -{\pi^4\over 24} (s^2+t^2+u^2) \str \l_a\l_b\l_c\l_d \cr
&+& {\pi^4\over 360} \left[ s(t-u) f_{abe}f_{cde}
+ t(s-u) f_{ace}f_{bde} + u(s-t) f_{ade}f_{bce} \right] \ .
\ea
Similarly at order $\ap^5$:
\ba\label{a45}
a_4^{(5)}&=& \pi^2 \zeta(3)\ s t u \, \str \l_a\l_b\l_c\l_d  \cr
&+& {c_4\over 48} (s^2+t^2+u^2) 
\left( t f_{abe}f_{cde} + s f_{ace}f_{bde}\right) \ . 
\ea
Finally, we give the order $\ap^6$ contribution to the four-point amplitudes:
\ba\label{a46}
a_4^{(6)}&=& - {\pi^6\over 480} (s^2+t^2+u^2)^2 \str \l_a\l_b\l_c\l_d  \cr
&+& {\pi^6\over 6048} (s^2+t^2+u^2) 
\left[ s(t-u)  f_{abe}f_{cde} + t(s-u) f_{ace}f_{bde}
+ u(s-t) f_{ade}f_{bce} \right] \cr
&-& \left( {\pi^6\over 3024} + {\zeta(3)^2\over 2}\right)stu
\left[  t f_{abe}f_{cde} + s f_{ace}f_{bde} \right]  \ . 
\ea
Note that, by construction, 
all $a_4^{(n)}$ are completely symmetric under exchange of any two 
external states, so that the symmetry properties of the amplitude are 
correctly given by those of the kinematical factors $K(1,2,3,4)$.

The explicit forms of the various four-point amplitudes
up to and including the order $\ap^2$ terms are given in \cite{BBdRS}.


\section{The effective action at order $\ap^2$}

In this section we set up the computation of the effective action. Since
the determination of the $\ap^3$ and $\ap^4$ terms in the next two sections
is an extension of the $\ap^2$ computation, it is 
most useful to first quickly review the latter as obtained in \cite{BBdRS} 
by matching the amplitudes computed from the 
effective action against the string amplitudes at order $\ap^2$. 
This is the purpose of the present section. 

\subsection{The $\ap$ expansion}

Our goal is to find the effective action which reproduces the $\ap$-expansion 
of the open superstring four-point amplitude of the previous section. 
Note that this determines the action only up to on-shell terms 
$\sim \g^\m \d_\m \c$ or $\sim \d_\m F^{\m\n}$.
At lowest order in $\ap$ this is of course well-known to be the ${\rm U}(N)$ 
${\cal N}=1$ super Yang-Mills theory in ten dimensions
\be\label{syn}
{\cal L}_{\rm SYM}
=\tr \left( -{1\over 4} F_{\m\n}F^{\m\n} +{i\over 2} \cb \g^\m D_\m \c\right).
\ee
The order $\ap^0$ amplitudes serve to fix the normalisations correctly. 
Our Feynman rules and other conventions are given in the appendix.
At higher orders in $\ap$, the
possible terms are given by dimensional analysis: in any space-time 
dimension $d$, the combination $g A_\m$ of the YM coupling constant and 
the gauge field has canonical dimension one, and thus $g F_{\m\n}$ has 
dimension two. Similarly, the analogous combination for the fermion $g \c$ 
has canonical dimension ${3\over 2}$. Thus dimensionless quantities are 
$\ap g F_{\m\n}$ and $\ap^2 g^2\, \cb \g D \c$ This leads to the following 
possible terms beyond ${\cal L}_{\rm SYM}$.

At order $\ap$ there is only $\ap g\, \cb \g D \c F$ while at order $\ap^2$ 
we have $(\ap g)^2\,  FFFF$, $(\ap g)^2\,  \cb \g D \c F F$ and 
$(\ap g)^2\,  \cb \g D \c\,  \cb \g D \c$ with various Lorentz and Lie algebra 
structures. One could also replace in some terms $g F$ by two covariant 
derivatives leading e.g. to $\ap^2 g \cb \g_\m D^\r D_\n D_\r \c F^{\m\n}$ or 
$\ap^2 g \cb \g^\r D_\m D_\r D_\n\c F^{\m\n}$ etc. Upon commuting two 
derivatives (which gives back $g F$) this contains either $D^2 \c$ or $\Ds\c$ 
which both vanish on-shell. Similarly, any term of the form $\ap^2 g F DF DF$ 
either vanishes on-shell ($\sim D_\r F^{\m\r}$), possibly after partial 
integration, or gives back some $(\ap g)^2 FFFF$ term. 
Thus there are no ``higher-derivative" terms at order $\ap^2$.

Note that all our order $\ap^2$ terms contain at least four fields 
($\c$ or $A_\m$) and hence each contribute to the four-point amplitude 
only via a single vertex diagram (1PI) while the order $\ap$ terms contain 
3, 4 or 5 fields and contribute to the four-point 
amplitude both a 1PI piece and via 
one-particle reducible $s$, $t$ and $u$-channel diagrams. All these terms 
would also contribute to higher-than-four-point, say six-point, amplitudes, 
both via one-particle reducible and irreducible diagrams.

At order $\ap^3$ we encounter
$(\ap g)^3 F^5$, but also $\ap^3 g^2 F^2 (DF)^2$ 
(which now does not vanish on-shell). 
Similarly at order $\ap^4$ we have e.g. $(\ap g)^4 F^6$ 
and $\ap^4 g^2 F^2 (D^2 F)^2$.
While the former terms cannot contribute to the four-point 
amplitude, the latter do. Comparing these contributions with the 
corresponding $\ap$-expansion of the string amplitude in
sections 4 and 5, we will determine
these higher order terms with no more 
than a total of four field strengths $F$ or fermion fields $\c$.

A remark is in order about on-shell terms. Consider e.g.
$\at^2 (\cb\ds\c)^2$ or $\at^2 \cb\ds\c F_{\r\s}F^{\r\s}$. They 
vanish on-shell and do not contribute to the {\it four}-point amplitude. 
These terms only contribute to 
the four-point amplitude via a single vertex (1PI diagram) 
and the fields involved are 
necessarily all on-shell. However, if we would compute the six-point 
amplitude at order $\ap^4$ using two such vertices joined by one 
fermion propagator in a 
one particle reducible diagram, the fermion might well be off-shell and 
such a term could not  be dropped. This is consistent with the following 
interpretation of on-shell terms: an order $\ap^2$ on-shell term can be 
removed by a field redefinition involving an order $\ap^2$ piece. But 
when substituting the new fields in the order $\ap^2$ interaction, this 
will also generate a new interaction at order $\ap^4$ that will contribute 
irreducibly to the six-point amplitude.

\subsection{The ansatz for the non-abelian effective action}

We write the effective action up to and including order $\ap^2$ as
\be\label{effaction}
{\cal L}={\cal L}_{\rm SYM} + {\cal L}_{\rm 4b}+ {\cal L}_{\rm 2b/2f}
+ {\cal L}_{\rm 4f} + {\cal L}_* +\O(\ap^3)
\ee
with ${\cal L}_{\rm 4b}$, ${\cal L}_{\rm 2b/2f}$ and ${\cal L}_{\rm 4f}$
containing the order $\ap^2$ terms needed to reproduce the string 
amplitudes to this order. 
The piece ${\cal L}_*$ contains any terms  
$\sim \at^2 \Ds\c$, $\sim \at^2 D_\m \cb \g^\m$, $\sim \at^2 D_\m F^{\m\n}$ 
that vanish on-shell and do not contribute to the four-point amplitudes 
as discussed above. In the following we write ${\cal L}_1 \simeq {\cal L}_2$ 
if  ${\cal L}_1$ and ${\cal L}_2$ only differ up to terms in ${\cal L}_*$ 
and up to partial integration.

The \underline{purely bosonic piece} ${\cal L}_{\rm 4b}$ is well-known 
since long \cite{GW}:
\be\label{fourbosl}
{\cal L}_{\rm 4b} = \at^2 \str \left( 
{1\over 8} F_{\m\n}F^{\n\r}F_{\r\s}F^{\s\m} 
-{1\over 32} \left( F_{\m\n}F^{\m\n}\right)^2 \right) 
\ee
where 
\be\label{alphatilde}
\at=2 \pi g \ap \ .
\ee
Since this contains exactly four $F$'s the 
contribution to the four gluon amplitude is obtained by extracting the 
interaction where each $F^a_{\m\n}$ is replaced simply by 
$\d_\m A_\n^a-\d_\n A_\m^a$. There is then a single order $\ap^2$ four 
gluon vertex contributing to the amplitude, and it is a straightforward 
exercise to show that the result coincides with the order $\ap^2$ part 
of the string amplitude $A_4^{\rm 4b}$.
In fact, it is not necessary to check all the terms in 
$K(\e_1,\e_2,\e_3,\e_4)$ since the structure of $K$ is fixed by gauge 
invariance and permutation symmetry. It is e.g. enough to check that 
\eqn{fourbosl} yields the 
$\e_1\cdot\e_2\, \e_3\cdot\e_4$ term  with the correct coefficient.
It is also easy to check that \eqn{fourbosl} is the unique interaction 
that reproduces the string four-gluon amplitude at order $\ap^2$.

The other terms in \eqn{effaction} were determined in \cite{BBdRS}. 
A possible order $\ap$ term 
$i c_1 d_{abc} \cb^a\g_\m D_\n\c^b F^{c\m\n}$ is first removed by a 
field redefinition 
$\c^a\to \c^a +{1\over 2} c_1 \ap d_{abc} F^b_{\r\s} \g_{\r\s} \c^c$. 
Then a general ansatz for the mixed piece is
\ba\label{bosfermlagrter}
{\cal L'}_{\rm 2b/2f}&=&
i \at^2 y_{abcd} \cb^a \g_\m D_\n \c^b F^{c\m\r}F^{d\ \ \n}_{\ \r}
+ i \at^2 z_{abcd} \cb^a\g_\m\g_\n\g_\r  D_\s \c^b F^{c\m\n}F^{d\r\s}\cr
&=&
i \at^2 {\widetilde y}_{abcd} \cb^a \g_\m D_\n \c^b F^{c\m\r}F^{d\ \ \n}_{\ \r}
+ i \at^2 z_{abcd} \cb^a\g_{\m\n\r}  D_\s \c^b F^{c\m\n}F^{d\r\s}
\ea
with
\be\label{ytilde}
{\widetilde y}_{abcd}=y_{abcd}+ 2 z_{abcd} 
\ee
and where the prime on ${\cal L}$ is just to remind us that this is the form 
after the field redefinition. 

For the \underline{four fermion interaction} ${\cal L}_{\rm 4f}$ a general
ansatz is
\be\label{lfourferm}
{\cal L}_{\rm 4f} = \at^2 g_{abcd} \cb^a \g^\m D^\n \c^b\, \cb^c \g_\m D_\n \c^d
+ \at^2 h_{abcd} \cb^a \g^\m D^\n \c^b\, \cb^c \g_\n D_\m \c^d \ ,
\ee
since other terms like
$\at^2 j_{abcd} \cb^a \g^{\m\n\r} D_\r \c^b\, \cb^c \g_\m D_\n \c^d$
or
$\at^2 l_{abcd} \cb^a \g^{\m\n\r} D^\s \c^b\, \cb^c \g_{\m\n\r} D_\s \c^d$
can be rewritten, using Fierz identities, as a combination 
of the two terms in \eqn{lfourferm}, up to terms $\sim \Ds\c$ which do 
not contribute to the amplitude. 
Upon partial integration and dropping any terms that vanish on-shell, one 
sees that one may just as well assume that $h_{abcd}$ is symmetric under 
interchange of $a$ and $b$ or of $c$ and $d$. Similarly, we may assume 
\be\label{gsymprop}
g_{abcd}=g_{cdab} \quad {\rm and} \quad
g_{(ab)[cd]}=g_{[ab](cd)}=0 \quad \Rightarrow \quad g_{abcd}=g_{badc} \ .
\ee

\subsection{Matching the 2 boson / 2 fermion amplitude}

The most convenient form of the 
relevant interaction is the first line of \eqn{bosfermlagrter}. It
only contributes two terms to the 2 boson / 2 fermion interaction, 
obtained upon replacing 
$D_\l\to \d_\l$ and $F^a_{\m\n}\to \d_\m A^a_\n - \d_\n A^a_\m$. 
Obviously, there is no order $\a'$ piece, while the computation of 
the order $\ap^2$ contribution to the amplitude is a  bit lengthy 
but straightforward. It yields \cite{BBdRS}
\ba\label{qftbosfermampl}
A_4^{\rm 2b/2f}&=& i \at^2 \Bigg\{ A( t z^++s z^-)\cr
& & \phantom{i\at^2} + \ub_1 \es_3 u_4 
\left[ 2 k_1\cdot \e_2 ( t z^+ +s z^-) 
+ {1\over 2} (t k_1\cdot \e_2 - s k_4\cdot \e_2) (y_{dacb}+y_{adcb}) \right] \cr
& & \phantom{i\at^2} - \ub_1 \es_2 u_4 
\left[ 2 k_4\cdot \e_3 ( t z^+ +s z^-) 
+ {1\over 2} (t k_4\cdot \e_3 - s k_1\cdot \e_3) (y_{dabc}+y_{adbc}) \right] \cr
& & \phantom{i\at^2} - \ub_1 \ks_3 u_4  \e_2\cdot \e_3
\left[ -2 t z^+ +{s\over 2} (y_{adbc}+y_{dacb}) 
- {t\over 2} (y_{dabc}+y_{adcb})\right] \cr
& & \phantom{i\at^2} + \ub_1 \ks_3 u_4 
\Big[ k_1\cdot \e_2 k_1 \cdot \e_3 (y_{dabc}-y_{dacb})
+  k_4\cdot \e_2 k_4 \cdot \e_3 (y_{adcb}-y_{adbc}) \cr
& & \phantom{i\at^2+ \ub_1 \ks_3 u_4}
- k_1\cdot \e_2 k_4 \cdot \e_3 (4z^- +y_{adbc}+y_{dacb})\cr
& & \phantom{i\at^2+ \ub_1 \ks_3 u_4}
+ k_4\cdot \e_2 k_1 \cdot \e_3 (4z^+ +y_{dabc}+y_{adcb}) \Big] \Bigg\}
\ea
with
\be\label{zplus}
z^+=z_{dabc}+z_{adcb} \quad , \quad z^-=z_{adbc}+z_{dacb}
\ee
and where $A$ (and $B$) where defined in \eqn{AB}. Note 
that this vanishes if we replace $\e_i\to k_i$, as required by gauge 
invariance.

Matching the result \eqn{qftbosfermampl} to the corresponding string 
amplitude  (recall that $\at=2\pi g \ap$)
yielded the following conditions \cite{BBdRS}
\be\label{matching}
z^+=z^- = -{1\over 4} \str \l_a\l_b\l_c\l_d \quad , \quad
y_{adbc}+y_{dabc} =  \str \l_a\l_b\l_c\l_d \quad , \quad
y_{adbc}= y_{adcb} \ .
\ee
This can be equivalently written as
\be\label{matchingbis}
z_{abcd}+z_{badc}= -{1\over 4} \str \l_a\l_b\l_c\l_d \quad , \quad 
y_{(ab)cd} =  {1\over 2} \str \l_a\l_b\l_c\l_d  \quad , \quad 
y_{ab[cd]}=0 \ .
\ee
It was shown in \cite{BBdRS} that, up to terms that secretly vanish on shell 
and hence can be eliminated by field redefinitions, the unique solution is
\ba\label{matchsol}
y_{abcd} &=& {1\over 2}  \str \l_a\l_b\l_c\l_d \cr
z_{abcd} &=& -{1\over 8} \str \l_a\l_b\l_c\l_d 
\ea

\subsection{Matching the 4 fermion amplitude}

We will now review the contribution of ${\cal L}_{\rm 4f}$ to the 
4 fermion amplitude. It was shown in \cite{BBdRS} that the second term in 
${\cal L}_{\rm 4f}$ cannot reproduce anything that looks like the 
string amplitude unless it can be transformed - using some Fierz 
identity - into a term with the same Lorentz index structure as  
the first one in 
${\cal L}_{\rm 4f}$. This is only possible if $h_{abcd}$ is completely symmetric in all its indices.
We begin by examining the contribution of the 
first term alone.  Obviously, its contribution to 
the amplitude contains $ \ub_1 \g_\m u_2 \ub_3 \g^\m u_4$,
$ \ub_1 \g_\m u_4 \ub_2 \g^\m u_3$ and 
$ \ub_1 \g_\m u_3 \ub_2 \g^\m u_4$. Using the Fierz identity
\eqn{fierz1} this last expression can be rewritten as a combination 
of the two other, and, upon taking into account \eqn{gsymprop} one
gets \cite{BBdRS}
\ba\label{a44fg}
A_4^{\rm 4f}\vert_{g-{\rm terms}}
&=&-2 i\at^2 \Bigg\{ 
\left[ (g_{acbd}+g_{adbc})\, s-g_{adcb}\, t-g_{acdb}\, u\right]
\ub_1 \g_\m u_4 \ub_2 \g^\m u_3 \cr
& & \phantom{-2 i\at^2 \Bigg\{   }
-\left[ (g_{acdb}+g_{abdc})\, u-g_{abcd}\, t-g_{acbd}\, s\right]
\ub_1 \g_\m u_2 \ub_3 \g^\m u_4 \Bigg\} \ .
\ea
Comparing with the string amplitude one sees that, if and only if
\be\label{gcond}
g_{abcd}=g_{acbd} \ ,
\ee
the amplitude reduces to the desired form
\be\label{a44fgbis}
A_4^{\rm 4f}\vert_{g-{\rm terms}}
=-2 i\at^2 \left( g_{acbd}+g_{abdc}+g_{adcb}\right)
\left( s\ \ub_1 \g_\m u_4 \ub_2 \g^\m u_3 
- u\ \ub_1 \g_\m u_2 \ub_3 \g^\m u_4 \right) \ .
\ee
The condition \eqn{gcond} together with \eqn{gsymprop} and the results of 
the appendix on 4-index tensors determine $g_{abcd}$ to be of the form
(dropping a piece that leads to a term that secretly vanishes on-shell)
\be\label{gform}
g_{abcd}=g \str \l_a\l_b\l_c\l_d
\ee
Concerning the second term it was shown \cite{BBdRS} that one needs
\be\label{hident}
h_{abcd} = h \str \l_a\l_b\l_c\l_d
\ee
so that one has using Fierz identities
\be\label{fierzfinal}
h_{abcd}\cb^a\g_\m D_\n\c^b \  \cb^c\g^\n D^\m\c^d
\simeq {2\over 3} h_{abcd} \cb^a\g_\m D_\n\c^b\  \cb^c \g^\m D^\n\c^d
= {2\over 3} h \str \cb\g_\m D_\n\c\  \cb \g^\m D^\n\c \ ,
\ee
and the $h$-term contributes to the amplitude ${2\over 3}$ of the 
$g$-term. Matching to the string amplitude then requires 
$3 g + 2 h = - {1\over 8}$.
While the expansion of the Born-Infeld determinant leads to 
$g={1\over 8}$ and $h=-{1\over 4}$, one should keep in mind that the 
$g$-term and  the $h$-term are really indistinguishable on-shell
(modulo the factor ${2\over 3}$), i.e. up to field redefinitions.

\subsection{The string effective action }

The full effective 
action up to and including all order $\ap^2\, g^2$ terms, bosonic, 
fermionic and mixed, can be written as \cite{BBdRS}
\ba\label{fulleffaction}
{\cal L}_{\rm string} &=& \str \Bigg( 
- {1\over 4} F_{\m\n}F^{\m\n} +{i\over 2} \cb \g^\m D_\m \c
+ {\at^2\over 8} F_{\m\n}F^{\n\r}F_{\r\s}F^{\s\m} 
- {\at^2\over 32} \left( F_{\m\n}F^{\m\n}\right)^2\cr
&&\phantom{\str\Bigg(} 
+i{\at^2\over 4} \cb \g_\m D_\n \c F^{\m\r} F_\r^{\ \n} 
-i{\at^2\over 8} \cb \g_{\m\n\r} D_\s \c F^{\m\n} F^{\r\s}\cr
&&\phantom{\str\Bigg(} 
+ {\at^2\over 8} \cb\g^\m D^\n\c\, \cb\g_\m D_\n \c
- {\at^2\over 4} \cb\g^\m D^\n\c\, \cb\g_\n D_\m \c \Bigg) 
+ {\cal O}(\at^3\, g^2\, , \ap^2\, g^3) \ .
\ea
As noted in \cite{BBdRS}, this coincides with the result of the following 
manipulation:
Take the abelian Born-Infeld action and expand it up to and including 
order $\ap^2$. Make the field redefinition to eliminate the order $\ap$ 
term, and drop all ``on-shell" terms $\sim \at^2 \ds\c$. 
Only then proceed to the obvious non-abelian generalisation 
and take a symmetrised trace.

\section{The effective action at order $\ap^3$}

We can now go on and compare results for the amplitudes beyond order $\ap^2$.
For the string amplitudes the $\ap$ expansion was easy to obtain and is given 
in section 2. As already discussed,  these four-point 
amplitudes only allow us to obtain information on the field theory side 
on terms  that are quartic 
(or less) in the fields. For example, at order $\ap^3$ we can get information 
about terms like $\ap^3 g^2 F^2 (D F)^2 $ but not $\ap^3 g^3 F^5$. Thus we 
cannot go beyond order $g^2$ terms.

We begin with the action \eqn{fulleffaction}. There are cubic and quartic 
vertices of order $\ap^0$ and quartic and higher vertices of order $\ap^2$. 
As a consequence there cannot be any contribution to the four-point 
amplitudes at order $\ap^3$ from  one-particle reducible diagrams with 
vertices made from the interactions already contained in  \eqn{fulleffaction}. 
All order $\ap^3$ contributions to the four-point amplitudes come from new 
quartic (and higher) interactions of order $\ap^3$. This same discussion 
then repeats itself at order $\ap^4$ and so on. This is a considerable 
simplification since disentangling one-particle reducible from one-particle 
irreducible contributions in general is a rather cumbersome task.

From eqs. \eqn{A4exp} and \eqn{a43} we know 
that the four-point string amplitudes at order $\ap^3$ are
\ba\label{A4alpha3}
A_4\vert^{\rm string}_{\ap^3}
&=&-8ig^2 \ap^3\ \zeta(3)\ \left( t f_{abe}f_{cde} + s f_{ace}f_{bde}\right)
K(1,2,3,4)\cr
&=&+8ig^2 \ap^3\ \zeta(3)\ \left( t f_{ade}f_{bce} + u f_{ace}f_{bde}\right)
K(1,2,3,4)    \ .
\ea

\subsection{Four fermion terms}

Lets first look at the four fermion interaction 
which we parametrize in a 
similar way as before:
\be\label{fourfermap3}
{\cal L}_{\rm 4f}^{\ap^3} =  \ct g^2 \ap^3 \gt_{abcd} \
\cb^a \g^\m D^{(\n} D^{\r)} \c^b\ \cb^c \g_\m D_{(\n} D_{\r)} \c^d 
\ee
with $\gt_{abcd}=\gt_{cdab}$. We wrote $D_{(\n} D_{\r)}$ since the antisymmetric
piece is $D_{[\n} D_{\r]}\sim g F_{\n\r}$ and corresponds to an order $g^3$ 
term which would show up in a five-point amplitude. Said differently, for our 
purpose of computing four-point amplitudes we can replace 
$D_\n D_\r \to \d_\n \d_\r$ which is symmetric in $\n$ and $\r$.
The computation of the amplitude then is 
very similar to the order $\ap^2$ computation above in \eqn{a44fg} except 
that  we have more derivatives in the interaction, and a priori, $\gt$ 
has less symmetries. The result is
\ba\label{a44fgt}
A_4^{\rm 4f}\vert_{\ap^3}
&=& {i\over 2}  \ct g^2 \ap^3 \Bigg\{ 
\Big[ (\gt_{abcd}+\gt_{badc}) t^2+(\gt_{acbd}+\gt_{cadb}) s^2 \cr
& & \phantom{-2 i\ct g^2\at^2 \Bigg\{ \Big[  }
-(\gt_{bacd}+\gt_{abdc}+\gt_{acdb}+\gt_{cabd}) u^2 \Big]
\ub_1 \g_\m u_2 \ub_3 \g^\m u_4 \cr
& & \phantom{-2 i\ct g^2\at^2   }
+\Big[ (\gt_{acbd}+\gt_{cadb}+\gt_{adbc}+\gt_{dacb}) s^2 \cr
& & \phantom{-2 i\ct g^2\at^2 \Bigg\{ \Big[  }
-(\gt_{acdb}+\gt_{cabd}) u^2 - (\gt_{adcb}+\gt_{dabc}) t^2 \Big]
\ub_1 \g_\m u_4 \ub_2 \g^\m u_3 \Bigg\} \ .
\ea
This can be matched to the string amplitude \eqn{A4alpha3} with
$K(u_1,u_2,u_3,u_4) = - {u\over 8}
\ \ub_1 \g_\m u_2 \ub_3 \g^\m u_4  
+{s\over 8}\ \ub_1 \g_\m u_4 \ub_2 \g^\m u_3$
if and only if the coefficients $\gt_{abcd}$ satisfy
\be\label{gtcond}
\gt_{abcd}+\gt_{badc}=2 f_{ade} f_{bce}
\ee
up to an arbitrary normalisation which we can absorb into $\ct$. Keeping 
in mind that $\gt_{abcd}=\gt_{cdab}$, the
general solution of this
condition is
\be\label{gtsol}
\gt_{abcd}=f_{ade} f_{bce} + \gt_4 (d_{abe} f_{cde}+d_{cde} f_{abe} )
\ee
with an arbitrary constant $\gt_4$.
Again, the pieces $\sim \gt_4$ in \eqn{fourfermap3} vanish on-shell and 
can be eliminated by a field redefinition. Hence we can assume $\gt_4=0$. 
The amplitude then reads
\be\label{a44fgtbis}
A_4^{\rm 4f}\vert_{\ap^3}
=2i \ct g^2 \ap^3 \left( t f_{abe}f_{cde} + s f_{ace}f_{bde}\right)
\left( u\ \ub_1\g_\m u_2\ub_3 \g^\m u_4  - s\ \ub_1\g_\m u_4\ub_2 \g^\m u_3
\right) \ .
\ee
Comparing with the string amplitude \eqn{A4alpha3} we find perfect agreement
if the constant $\ct$ is choosen to be $\ct={1\over 2} \zeta(3)$.
Then the four fermion interaction at order $\ap^3$ reads
\be\label{fourfermap3final}
{\cal L}_{\rm 4f}^{\ap^3} = {\zeta(3)\over 2}   g^2 \ap^3 f_{ade}f_{bce}
\cb^a \g^\m D^\n D^\r \c^b\ \cb^c \g_\m D_\n D_\r \c^d \ .
\ee
Note that, as always with four fermion terms, 
${\cal L}_{\rm 4f}^{\ap^3}$  can be rewritten in a variety of ways using 
the Fierz transformations. It is clear
nevertheless that there is no way to rewrite it as a symmetrised trace. 
Thus: {\it there is no symmetrised trace prescription 
at order $\ap^3$ }! No field redefinition or Fierz transformation
could help to evade this conclusion.

\subsection{Two boson / two fermion interaction}

In analogy with the order
$\ap^2$ interaction  ${\cal L'}_{\rm 2b/2f}$ of \eqn{bosfermlagrter} we
start with
\ba\label{bosfermap3}
{\cal L}_{\rm 2b/2f}^{\ap^3}=
i g^2 \ap^3 & \Big(&
Y^{(1)}_{abcd}\ \cb^a \g_\m D_\n D_\l \c^b\, F^{c\m\r} D_\l F^{d\ \ \n}_{\ \r}
+ Y^{(2)}_{abcd}\ \cb^a \g_\m D_\n D_\l \c^b\, D_\l F^{c\m\r} F^{d\ \ \n}_{\ \r}
\cr
&+& Z^{(1)}_{abcd}\ \cb^a \g_\m \g_\n \g_\r D_\s D_\l \c^b\, 
F^{c\m\n} D_\l F^{d\r\s}
+ Z^{(2)}_{abcd}\ \cb^a \g_\m \g_\n \g_\r D_\s D_\l \c^b\, 
D_\l F^{c\m\n} F^{d\r\s}  \Big) \cr
& &
\ea
Again we may assume $D_\n D_\l \to D_{(\n} D_{\l)}$ etc.
With respect to eq.  \eqn{bosfermlagrter} we have two more derivatives and
$y_{abcd}$ is replaced by $Y^{(1)}_{abcd}$ or $Y^{(2)}_{abcd}$ and
$z_{abcd}$ by $Z^{(1)}_{abcd}$ or $Z^{(2)}_{abcd}$. It is easy to see that the
amplitude computation then proceeds in exactly the same way, except that extra
factors of $s$, $t$ or $u$ appear. One can copy these computations line by line
if one makes the following substitutions (in addition to $\at^2\to g^2 \ap^3$):
\ba\label{substi}
y_{adbc} &\to& - {s\over 2} Y^{(1)}_{adbc} -{t\over 2} Y^{(2)}_{adbc} \ , \quad
y_{adcb}\to -{t\over 2} Y^{(1)}_{adcb}-{s\over 2} Y^{(2)}_{adcb} \cr
y_{dabc} &\to& - {t\over 2} Y^{(1)}_{dabc} -{s\over 2} Y^{(2)}_{dabc}  \ , \quad
y_{dacb}\to -{s\over 2} Y^{(1)}_{dacb}-{t\over 2} Y^{(2)}_{dacb} \cr
z^+ &\to& -{t\over 2} Z^{(1)+} - {s\over 2}Z^{(2)+} \ , \quad
z^- \to -{s\over 2} Z^{(1)-} - {t\over 2}Z^{(2)-}
\ea
where $Z^{(1)\pm}$ and $Z^{(2)\pm}$ are defined in analogy with \eqn{zplus} 
for $z^\pm$. We perform these substitutions in the resulting amplitude
\eqn{qftbosfermampl} and match the resulting expression to the string amplitude
$\sim uA+sB$. Vanishing of the $ \ub_1 \ks_3 u_4\  k\cdot\e\, k\cdot\e$ terms 
requires
\be\label{yoneytwo}
Y^{(2)}_{abcd}=Y^{(1)}_{abdc}
\ee
and
\be\label{zycond}
{\cal Z}\equiv s Z^{(1)-}+t Z^{(2)-} = t Z^{(1)+}+s Z^{(2)+}
=-{s\over 4} \left( Y^{(1)}_{adbc}+Y^{(1)}_{dacb} \right)
-{t\over 4} \left( Y^{(1)}_{adcb}+Y^{(1)}_{dabc} \right) \ .
\ee
Then the amplitude becomes
\be\label{fermbosampa3}
A_4^{\rm 2b/2f} \vert_{\ap^3} = {i\over 2} g^2 \ap^3 (uA+sB)\ {\cal Z}  = 
4i g^2 \ap^3 K(u_1,\e_2,\e_3,u_4)\  {\cal Z}
\ee
so that we need
\be\label{zdeterm}
{\cal Z}=- 2 \zeta(3)  (t f_{abe}f_{cde}+s f_{ace}f_{bde} )\ .
\ee
Then the general solution of eq. \eqn{zycond} is 
\ba\label{zysolutions}
Y^{(1)}_{abcd} = -4\zeta(3)  f_{ade}f_{bce} 
+ y_4 d_{abe}f_{cde} + y_5 d_{cde}f_{abe} \ , \cr
Y^{(2)}_{abcd} = -4\zeta(3)  f_{ace}f_{bde} 
- y_4 d_{abe}f_{cde} + y_5 d_{cde}f_{abe}\ , \cr
Z^{(1)}_{abcd} = \zeta(3)  f_{ade}f_{bce}
+ z_4^{(1)} d_{abe}f_{cde} + z_5^{(1)} d_{cde}f_{abe}\ , \cr
Z^{(2)}_{abcd} = \zeta(3)  f_{ace}f_{bde}
+ z_4^{(2)} d_{ace}f_{bde} + z_5^{(2)} d_{bde}f_{ace}\  .
\ea
As before, there are undetermined parameters $y_4, y_5, z_4^{(1)},
z_5^{(1)}, z_4^{(2)}$ and $z_5^{(2)}$ but the corresponding terms in 
\eqn{bosfermap3} all vanish on-shell which can be shown along the same 
lines as in \cite{BBdRS}. They can thus be eliminated by field 
redefinitions and we can assume 
$y_4= y_5= z_4^{(1)}=z_5^{(1)}= z_4^{(2)}=z_5^{(2)}=0$ from the outset.
The interaction \eqn{bosfermap3} then takes the very simple form
\ba\label{bosfermap3bis}
{\cal L}_{\rm 2b/2f}^{\ap^3}=2 i \zeta(3)\, g^2 \ap^3 f_{ade} f_{bce} & &
\Big\{ -4 \cb^a \g_{(\m} D_{\n)} D_\l \c^b\ F^{c\m\r} D_\l F^{d\ \n}_{\ \r} \cr
& & + \cb^a \g_\m \g_\n \g_\r D_\s D_\l \c^b\ {1\over 2}
\left( F^{c\m\n} D_\l F^{d\r\s} + F^{c\r\s} D_\l F^{d\m\n} \right) \Big\}
\ea
which can also be rewritten as
\be\label{bosfermap3ter}
{\cal L}_{\rm 2b/2f}^{\ap^3}= 2\zeta(3)\, g^2 \ap^3 f_{ade} f_{bce} 
\Big\{ -2 i \cb^a \g_{(\m} D_{\n)} D_\l \cb\ F^{c\m\r} D_\l F^{d\ \n}_{\ \r} 
+ i \cb^a \g_{\m\r [\s} D_{\n]}  D_\l \c^b \
F^{c\m\n} D_\l F^{d\r\s} \Big\} \ .
\ee
This last form is particularly suggestive when compared to the order $\ap^2$
interaction.

\subsection{Four boson interaction}

We start with the following
interaction
\ba\label{fourbosap3}
{\cal L}_{\rm 4b}^{\ap^3} = g^2 \ap^3    \Big( &&
\a_{abcd}\, F^{a}_{\m\n} D_\l F^{b\n\r} D_\l F^c_{\r\s} F^{d\s\m}
+\b_{abcd}\, F^{a}_{\m\n} D_\l F^{b\n\r}  F^c_{\r\s} D_\l F^{d\s\m} \cr
&&
+\g_{abcd}\,  F^a_{\m\n} D_\l F^{b\m\n}\, F^c_{\r\s} D_\l F^{d\r\s} \Big)
\ea
where obviously we may assume that the coefficients have the 
following symmetries:
$\a_{abcd}=\a_{dcba}$, $\b_{abcd}=\b_{cdab}=\b_{cbad}=\b_{adcb}$
and $\g_{abcd}=\g_{cdab}$. 
It is slighly less obvious and needs some partial integration, 
reshuffling of indices and dropping of on-shell terms, to show 
that we may also assume
$\a_{(ab)[cd]}=\a_{[ab](cd)} =\g_{(ab)[cd]}=\g_{[ab](cd)}=0$.
Using the results of the appendix, these symmetries imply that
\ba\label{abcdecomp}
\a_{abcd}&=&\a_1 d_{abe}d_{cde} + \a_2 d_{ace}d_{bde} + \a_3
d_{ade}d_{bce}\cr
\b_{abcd}&=&
\b_1 (d_{abe}d_{cde} + d_{ade}d_{bce}) + \b_2 d_{ace}d_{bde}
\cr
\g_{abcd}&=&
\g_1 d_{abe}d_{cde} + \g_2 d_{ace}d_{bde} + \g_3
d_{ade}d_{bce} \ .
\ea
Note that we have not written a term
$\sim\delta_{abcd} D_\l F^a_{\m\n} D_\l F^{b\m\n} F^c_{\r\s} F^{d\r\s}$
with a $\delta_{abcd}$ that is symmetric under exchange of $a$ and $b$, of
$c$ and $d$ and of $ab$ with $cd$, since upon partial integration it
can
be rewritten (on shell) as the symmetric part of the
term $\sim\g_{abcd}$.
As already discussed, it is not necessary to check that the full
string amplitude with all terms in
$K(\e_1,\e_2,\e_3,\e_4)$ is reproduced. Since all terms in $K$ are
uniquely determined from any single one in $K$ by permutation symmetry
and gauge invariance, it is enough to check e.g. that the term
$-{ut\over 4} \e_1\cdot\e_2\, \e_3\cdot\e_4$ is correctly reproduced
by the interaction. Thus we want to obtain a term
\be\label{epsepsterm}
-2i \zeta(3)\, g^2 \ap^3 ( t f_{ade}f_{bce} + u f_{ace} f_{bde} )\ u t \
\e_1\cdot\e_2\, \e_3\cdot\e_4 \ .
\ee
The interaction \eqn{fourbosap3} leads to
\ba\label{fourepsap3term}
-{i\over 4} g^2 \ap^3
\e_1\cdot\e_2\, \e_3\cdot\e_4  &\Big\{ 
 &t^3 (\a_{abdc}+a_{bacd}) + u^3 (\a_{abcd}+\a_{badc}) 
+su^2 (\a_{adcb}+\a_{dabc}) + st^2 (\a_{acdb}+\a_{cabd}) \cr
&+&2t^2u(\b_{abdc}+\b_{bacd}) +2
tu^2 (\b_{abcd}+\b_{badc}) \cr
&+&4s^2t(\g_{abcd}+\g_{badc}) +4s^2u(\g_{abdc}+\g_{bacd}) \Big\} \ .
\ea
This equals the string amplitude if and only if
\ba\label{abccond}
0&=& \a_{abdc}+\a_{badc}- (\a_{acdb}+\a_{cabd}) + 4(\g_{abcd}+\g_{badc})
\cr
-8\zeta(3)   f_{ade} f_{bce} &=&
-(\a_{acdb}+\a_{cabd}) + 2(\b_{abdc}+\b_{bacd})
  +8(\g_{abcd}+\g_{badc})+4(\g_{abdc}+\g_{bacd}) \ .
\ea
This is solved by
\ba\label{abcsol}
\a_{abcd}&=& -\zeta(3)  f_{ace} f_{bde} \cr
\b_{abcd}&=& - {\zeta(3)\over 2} (f_{ade}f_{bce}-f_{abe}f_{cde})\cr
\g_{abcd} &=& {\zeta(3)\over 2}  f_{ade}f_{bce}
\ea
While this is a solution, it is not the most general one. Indeed,
plugging in the general form \eqn{abcdecomp} of the coefficients into
the equations \eqn{abccond} yields
\be\label{abcsoltwo}
\b_1=-{\zeta(3)\over 2}+{\a_0\over 2}
\ , \ \ 
\b_2=\zeta(3) +{\a_2\over 2}
 \ , \ \
\g_1=-\g_2={\zeta(3)\over 2}
 \ , \ \ 
\g_3=0 
\ , \ \
\a_3-\a_1= 2\zeta(3) 
\ee
with $\a_2$ and $\a_1+\a_3 \equiv 2\a_0$ undetermined. 
The general solution then is
\ba\label{abcgensol}
\a_{abcd}&=& -\zeta(3) f_{ace} f_{bde} 
+ \a_0 ( d_{abe}d_{cde}+d_{ade}d_{bce}) + \a_2 d_{ace}d_{bde}\cr
\b_{abcd}&=& -{\zeta(3)\over 2}(f_{ade}f_{bce}-f_{abe}f_{cde})
+ {\a_0\over 2} (d_{abe}d_{cde}+d_{ade}d_{bce}) 
+{\a_2\over 2}d_{ace}d_{bde}\cr
\g_{abcd} &=&  {\zeta(3)\over 2} f_{ade}f_{bce} \ .
\ea
Again, the terms involving the ambigous $\a_0$ and $\a_2$ actually vanish. 
The $\a_2$ term in $\a_{abcd}$ e.g. leads to a term 
$\a_2 d_{ace}d_{bde} F^a_{\m\n} D_\l F^{b\n\r} D_\l F^c_{\r\s} F^{d\s\m}
={\a_2\over 2} D_\l (d_{ace} F^a_{\m\n} F^c_{\r\s}) 
d_{bde} D_\l F^{b\n\r} F^{d\s\m}$ which, upon partial integration cancels 
the term coming from the $\a_2$ contribution in $\b_{abcd}$. Things work 
out similarly for the $\a_0$ terms.
Hence we can set $\a_0=\a_2=0$ without loss of generality and the solution
is uniquely given by \eqn{abcsol} up to field redefinitions. The 
interaction \eqn{fourbosap3}
then takes the following form
\be\label{fourbosap3final}
{\cal L}_{\rm 4b}^{\ap^3} = - 2\zeta(3) g^2 \ap^3 f_{ade} f_{bce} 
\left\{ {1\over 2} F^a_{\m\n} D_\l F^{b\n}_{\ \ \r}
\left( F^{c\r}_{\ \ \s}  D_\l F^{d\s\m} 
+ F^{c\m}_{\ \ \s} D_\l F^{d\s\r} \right)
-{1\over 4} F^a_{\m\n} D_\l F^{b\m\n} F^c_{\r\s} D_\l F^{d\r\s}
\right\} \ .
\ee

\subsection{ The string effective action at order $\ap^3$}

One should keep in mind that at order $\ap^3$  one can also
write other terms with less derivatives and more fields, like $g^3 \ap^3
f_{abf} f_{cdg} f_{fge} F^{a\m}_{\ \ \n} F^{b\n}_{\ \ \r} F^{c\r}_{\ \
\s} F^{d\s}_{\ \ \l} F^{e\l}_{\ \ \m}$ or mixed terms involving two
fermion fields and three $F$'s or four fermion fields and one $F$. All
these terms are of order $g^3$ and involve at least five fields so that
they will only show up when computing five-point amplitudes. 
We have nothing to say about them here.

We now summarise all order $\ap^3\, g^2$ terms we have extracted 
from the string four-point amplitude.
This higher-derivative piece of the string effective action is uniquely 
determined (up to field redefinitions) as
\ba\label{fullap3effaction}
{\cal L}_{\rm 4\ fields}^{\ap^3} &=&
2 \zeta(3)\, g^2 \ap^3\, f_{ade} f_{bce} \ \times \cr
&\Big\{&
-{1\over 2} F^a_{\m\n} D_\l F^{b\n}_{\ \ \r}
\left( F^{c\r}_{\ \ \s}  D_\l F^{d\s\m} 
+ F^{c\m}_{\ \ \s} D_\l F^{d\s\r} \right)
+{1\over 4} F^a_{\m\n} D_\l F^{b\m\n} F^c_{\r\s} D_\l F^{d\r\s}\cr
&-&
  2 i \cb^a \g_{(\m} D_{\n)} D_\l \cb\ F^{c\m\r} D_\l F^{d\ \n}_{\ \r} 
+ i \cb^a \g_{\m\r [\s} D_{\n]}  D_\l \c^b \
F^{c\m\n} D_\l F^{d\r\s} \cr
&+&
{1\over 4}
\cb^a \g^\m D^\n D^\r \c^b\ \cb^c \g_\m D_\n D_\r \c^d \Big\}
\ea

\section{The effective action at order $\ap^4$}

It  is not difficult to continue this exercise at order $\ap^4$. The string 
amplitudes we want to reproduce  are (cf \eqn{a44})
\ba\label{A4ap4}
A_4\vert^{\rm string}_{\ap^4}
= {i\over 3} \pi^4 g^2 \ap^4 &\Big\{&
(s^2+t^2+u^2) \str \l_a\l_b\l_c\l_d \
- \ {1\over 15} \Big[ s(t-u) f_{abe}f_{cde} \cr
&&+ t(s-u) f_{ace}f_{bde} + u(s-t) f_{ade}f_{bce} \Big]\ \Big\}\ 
K(1,2,3,4) \ .
\ea
They now contain both, a symmetrised trace piece and an $f\ f$ piece.

\subsection{Four fermions}

We start with an interaction similar to 
\eqn{fourfermap3}:
\be\label{fourfermap41}
{\cal L}_{\rm 4f, 1}^{\ap^4} =  \ch_1\, g^2 \ap^4\, \gh_{abcd} \
\cb^a \g^\m D^\n D^\r D^\s \c^b\ \cb^c \g_\m D_\n D_\r D_\s \c^d 
\ee
with $\gh_{abcd}=\gh_{cdab}$ where again we can replace $D\to \d$ at the 
order $g^2$ we are working. We can then copy the calculation leading to
\eqn{a44fgt}, except that now $\ct g^2\ap^3\to \ch_1\, g^2\ap^4$, 
$\gt_{....}\to\gh_{....}$ and the extra derivatives lead to the replacements
\be\label{stureplace}
t^2\to -{t^3\over 2} \ , \quad
u^2\to -{u^3\over 2} \ , \quad
s^2\to -{s^3\over 2} \ .
\ee
Requiring that this amplitude be proportional to $K(u_1,u_2,u_3,u_4)$ implies
\be\label{ghatcond}
\gh_{abcd}+\gh_{badc}=\gh_{adbc}+\gh_{dacb}=\gh_{adcb}+\gh_{dabc}
\ee
so that the contribution to the amplitude then is
\be\label{fourfermcontr1}
A_4^{\rm 4f}\vert_{\ap^4}^{(1)} 
= - 3 i \ch_1\,  g^2 \ap^4\,  (\gh_{abcd}+\gh_{badc}) 
(s^2+t^2+u^2) K(u_1,u_2,u_3,u_4) \ .
\ee
We see that ${\cal L}_{\rm 4f}^{\ap^4}$ can only reproduce the symmetrised 
trace part of the string amplitude, and it does so correctly provided
\be\label{ghatconehatcond}
\gh_{abcd}=\str \l_a\l_b\l_c\l_d \ , \quad \ch_1=-{\pi^4\over 18} \ .
\ee

To reproduce the $f\ f$ piece, we need a different interaction. An
interaction of the form 
$\sim D^\n\cb^a \g^\m  D^\r D^\s \c^b\ D_\n\cb^c \g_\m  D_\r D_\s \c^d$
does not help since it again leads to \eqn{stureplace} and hence to 
\eqn{fourfermcontr1}. If instead we start with
\be\label{fourfermap42}
{\cal L}_{\rm 4f, 2}^{\ap^4} =  \ch_2 g^2 \ap^4 \hh_{abcd} \
D^\n \cb^a \g^\m  D^\r D^\s \c^b\ D_\r \cb^c \g_\m D_\n D_\s \c^d 
\ee
we get
\ba\label{fourfermcontr2}
A_4^{\rm 4f}\vert_{\ap^4}^{(2)} = -{i\over 4} \ch_2 g^2 \ap^4
&\Big\{&  \Big[
-(\hh_{abcd}+\hh_{badc}) ut - (\hh_{acbd}+\hh_{cadb}) us \cr
&&+(\hh_{bacd}+\hh_{abdc})t^2+(\hh_{acdb}+\hh_{cabd})s^2 \Big]
(-u\, \ub_1\g^\m u_2\, \ub_3\g_\m u_4 ) \cr
&+& \Big[ (\hh_{acbd}+\hh_{cadb})u^2 + (\hh_{adbc}+\hh_{dacb})t^2\cr
&&-(\hh_{acdb}+\hh_{cabd})su - (\hh_{adcb}+\hh_{dabc})st \Big]
(s\, \ub_1\g^\m u_4\, \ub_2\g_\m u_3 ) \Big\} \ .
\ea
Matching this to the $f\ f$ part of the string amplitude requires 
(up to an overall normalisation)
\ba\label{hhatcond}
\hh_{abcd}+\hh_{badc} + \hh_{bacd}+\hh_{abdc} &=&
2f_{ace}f_{bde}+2f_{ade}f_{bce} \cr
\hh_{acbd}+\hh_{cadb} + \hh_{acdb}+\hh_{cabd} &=&
-2f_{ade}f_{bce}+2f_{abe}f_{cde} \cr
\hh_{bacd}+\hh_{abdc} + \hh_{acdb}+\hh_{cabd} &=&
-2f_{abe}f_{cde}-2f_{ace}f_{bde} \ .
\ea
This is solved by
\be\label{hhatsol}
\hh_{abcd}=f_{abe}f_{cde}+f_{ace}f_{bde}
\ee
and \eqn{fourfermcontr2} equals the $f\ f$ part of the string amplitude provided the normalisation is chosen as
\be\label{chat2cond}
\ch_2={\pi^4\over 180} \ .
\ee
The full four fermion interaction at order $\ap^4$ then is
\ba\label{fourfermap4}
{\cal L}_{\rm 4f}^{\ap^4} = - {\pi^4\over 18} \, g^2 \ap^4
&\Big\{& \str \cb\g^\m D^\n D^\r D^\s\c\ \cb\g_\m D_\n D_\r D_\s\c \cr
&-&{1\over 10} (f_{abe}f_{cde}+f_{ace}f_{bde})
D^\n \cb^a \g^\m  D^\r D^\s \c^b\ D_\r \cb^c \g_\m D_\n D_\s \c^d \Big\} \ .
\ea

\subsection{Four bosons} 

It is straightforward to
extend the discussion to the other cases as well. Clearly there will 
be a symmetrised 
trace part and an $f\ f$ part in each case. Since two more derivatives have 
to be distributed than for the $\ap^3$ interaction, many more terms are 
possible in the interaction to start with. 

We begin by considering the 
symmetrised trace part of the four boson interaction
${\cal L}_{\rm 4b,\ sym\ trace\ part}^{\ap^4}$.
One can write down nine different 
tensor structures with the four derivatives and four field strengths. Four 
of these structures are related in an obvious way to the other by partial 
integration 
and up to on-shell terms $\sim D_\l D^\l F_{\m\n}$ that do not contribute 
to the amplitude, leaving a general ansatz for 
${\cal L}_{\rm 4b,\ sym\ trace\ part}^{\ap^4}$
with only five different structures. Thus a priori we have 
five coefficients to be determined.
As before it is enough to match a single term in $K(\e_1,\e_2,\e_3,\e_4)$ 
in the amplitudes, e.g. the term $- {ut\over 4} \e_1\cdot\e_2\, \e_3\cdot\e_4$.
In particular, one sees that, due to the factor $ut$, the part of the 
amplitude containing $\e_1\cdot\e_2\, \e_3\cdot\e_4$ may contain 
(after replacing any $s$ by $-t-u$)
$u^3t,\ u^2t^2$ or $ut^3$ but not $u^4$ or $t^4$. Vanishing of the $u^4+t^4$
terms imposes one relation between these coefficients.
Equality of the $ut(u^2+t^2)$ terms with the $ut\, ut$ terms, as needed 
to reproduce an overall factor $s^2+t^2+u^2$ gives another condition. 
Matching the overall normalisation then gives a third relation, so that
we are left with two undetermined parameters, say $a$ and $b$. The details of 
the computation are by now rather straightforward and we only give the result.
One finds
\ba\label{fourbosap4symtrace}
{\cal L}_{\rm 4b,\ sym\ trace\ part}^{\ap^4} = {\pi^4\over 3}\, g^2 \ap^4
\str &\Big\{& F_{\m\n} D^\l D^\k F^{\n\r}
\Big[ a D_\l D_\k F_{\r\s} F^{\s\m} + b F_{\r\s} D_\l D\k F^{\s\m}\cr
& &\phantom{F_{\m\n} D^\l D^\k F^{\n\r}\Big[ }
+(a+2b-1) D_\l F_{\r\s} D_\k F^{\s\m} \Big] 
\cr
&-& {1\over 4} F_{\m\n} D^\l D^\k F^{\m\n} \left[ 2a F_{\r\s} D_\l D_\k F^{\r\s}
+(1-a) D_\l F_{\r\s} D_\k F^{\r\s} \right]
\Big\}\cr
& &
\ea
with arbitrary parameters $a$ and $b$.
However, it is not too difficult to see that the terms $\sim b$ actually 
vanish on-shell, up to partial integration. The same is also true 
for the terms $\sim a$ but to show this is slightly more tricky and 
requires repeated use of the Bianchi identity. As a result, the 
$\str$-part of ${\cal L}_{\rm 4b}^{\ap^4} $ is uniquely determined 
and the choices of $a$ and $b$ are irrelevant at this level. Convenient 
choices may be $a=1,\ b=0$ or $a=b={1\over 3}$ or even $a=b=0$ 
which lead to somewhat 
more elegant forms of ${\cal L}_{\rm 4b,\ sym\ trace\ part}^{\ap^4}$
than do other choices.

To determine the part of the four boson interaction which involves the 
products of two structure constants, in analogy with \eqn{fourbosap4symtrace},
we take the following ansatz
\ba\label{fourbosap4ffpart}
{\cal L}_{{\rm 4b},\ f f\  {\rm part}}^{\ap^4} = {\pi^4 g^2 \ap^4\over 90} 
&\Big\{& 
h^{(1)}_{abcd} F^a_{\m\n} D^\l D^\k F^{b\n\r} 
D_\l D_\k F^c_{\r\s} F^{d\s\m} \cr
&+&h^{(2)}_{abcd} F^a_{\m\n} D^\l D^\k F^{b\n\r} 
F^c_{\r\s} D_\l D_\k F^{d\s\m} \cr
&+&h^{(3)}_{abcd} F^a_{\m\n} D^\l D^\k F^{b\n\r} 
D_\l F^c_{\r\s} D_\k F^{d\s\m} \cr
&+&h^{(4)}_{abcd} F^a_{\m\n} D^\l D^\k F^{b\m\n} 
F^c_{\r\s} D_\l D_\k F^{d\r\s} \cr
&+&h^{(5)}_{abcd} F^a_{\m\n} D^\l D^\k F^{b\m\n} 
D_\l F^c_{\r\s} D_\k F^{d\r\s}
\Big\} \ . 
\ea
We parametrise the coefficients $h^{(i)}_{abcd}$ as
\be\label{hiabcd}
h^{(i)}_{abcd}=\a_i f_{ace}f_{bde} + \b_i f_{ade}f_{bce}
\ee
As before it is enough to match a single term in $K(\e_1,\e_2,\e_3,\e_4)$ 
in the amplitudes, and we again consider 
$- {ut\over 4} \e_1\cdot\e_2\, \e_3\cdot\e_4$.
Again, the part of the 
amplitude containing $\e_1\cdot\e_2\, \e_3\cdot\e_4$ may contain 
$u^3t,\ u^2t^2$ or $ut^3$ but not $u^4$ or $t^4$.
Extracting all terms $\sim \e_1\cdot\e_2\, \e_3\cdot\e_4$ in the amplitude 
as computed from \eqn{fourbosap4ffpart} yields, among others, these unwanted
terms 
$\sim u^4$ or $\sim t^4$. They are
\ba\label{u4t4terms}
{\pi^4 g^2 \ap^4\over 1440}\e_1\cdot\e_2\, \e_3\cdot\e_4 &\Big\{&
u^4 \left[ 4(\b_1+4\a_4)f_{ade}f_{bce} +2(8\b_4-\b_1)f_{ace}f_{bde} \right] \cr
&+&t^4 \left[ 4(\b_1+4\a_4)f_{ace}f_{bde} +2(8\b_4-\b_1)f_{ade}f_{bce} \right] 
\Big\} 
\ea
and must vanish. This implies $\a_4=-2\b_4= - {1\over 4}\b_1$ so that
\be\label{h4abcd}
h^{(4)}_{abcd}=- \b_4 ( f_{ace}f_{bde} +  f_{abe}f_{cde} )
\ee
which is exactly the same combination of the $f$-tensors as appeared 
above in the corresponding part of the 4 fermion interaction. It is 
quite reasonable to guess that the same combination will also be 
determined for the other $h^{(i)}_{abcd}$. To somewhat simplify 
things, we will assume this from the outset. So we make this ansatz 
also for the other $h^{(i)}_{abcd}$:
\be\label{hiansatz}
h^{(i)}_{abcd}=- \b_i ( f_{ace}f_{bde} +  f_{abe}f_{cde} ) 
\quad, i=1, \ldots 5 \ .
\ee
Introducing the notation
\be\label{F1F2}
F_1=f_{ace}f_{bde} \ , \quad F_2=  f_{abe}f_{cde}
\ee
the part of the amplitude $\sim \e_1\cdot\e_2\, \e_3\cdot\e_4$ as 
obtained from \eqn{fourbosap4ffpart} then is
\ba\label{ffamplotherterms}
{\pi^4 g^2 \ap^4\over 720}\ \e_1\cdot\e_2\, \e_3\cdot\e_4 &\Big\{&
u^3t\left[
(2\b_1-3\b_3-16\b_5)F_1+(2\b_1+8\b_5)F_2
\right] \cr
&+&ut^3\left[
(2\b_1-3\b_3-16\b_5)F_1+(-4\b_1+3\b_3+8\b_5)F_2
\right] \cr
&+&u^2t^2\left[
(2\b_1+2\b_2-4\b_3-32\b_5)F_1+(-\b_1-\b_2+2\b_3+16\b_5)F_2
\right] \Big\}  \ .
\ea
This must equal the corresponding contribution in the string amplitude which is
\be\label{stringffbosonap4}
{\pi^4 g^2 \ap^4\over 720}\ \e_1\cdot\e_2\, \e_3\cdot\e_4 \Big\{
u^3t \left[ -4F_1+8F_2\right] + ut^3 \left[ -4F_1-4F_2\right]
+u^2t^2 \left[ -16F_1+8F_2\right] \Big\} \ ,
\ee
leading to six conditions, but only three of them are linearly independent. 
This allows us to solve for $\b_2, \b_3$ and $\b_5$ in terms of 
$\b_1\equiv 4 \b$. Then we have for all five coefficients the following 
parametrisation:
\be\label{betasol}
\b_1=4\b\ , \quad \b_2=-4\b \ , \quad \b_3=8\b-4\ , \quad 
\b_4={1\over 2}\b \ , \quad \b_5=1-\b \ .
\ee
The interaction \eqn{fourbosap4ffpart} then takes the following form
\ba\label{fourbosap4ffpartsol}
{\cal L}_{{\rm 4b},\ f f\  {\rm part}}^{\ap^4} &=& - {2\pi^4 g^2 \ap^4\over 45}
( f_{ace}f_{bde} +  f_{abe}f_{cde} ) \times \cr
&\times&\Big\{ 
F^a_{\m\n} D^\l D^\k F^{b\n\r} 
\left[ \b D_\l D_\k F^c_{\r\s} F^{d\s\m} - 
\b F^c_{\r\s} D_\l D_\k F^{d\s\m} 
+(2\b-1) D_\l F^c_{\r\s} D_\k F^{d\s\m} \right] \cr
&&+ {1\over 4} F^a_{\m\n} D^\l D^\k F^{b\m\n} 
\left[ {\b\over 2} F^c_{\r\s} D_\l D_\k F^{d\r\s} 
+ (1-\b)  D_\l F^c_{\r\s} D_\k F^{d\r\s} \right] 
\Big\} \ .
\ea
The parameter $\b$ is arbitrary and one can probably show again that 
the expression multiplying $\b$ vanishes on-shell after partial 
integration and use of the Bianchi identity. A choice leading to a 
particularly simple interaction is $\b=0$.

Combining the symmetrised trace part for $a=b=0$ and the $f\, f$-part 
with $\b=0$ gives (one form of) the full 4 boson term at order $\ap^4$:
\ba\label{fourbosap4}
{\cal L}_{{\rm 4b}}^{\ap^4} = - {\pi^4 g^2 \ap^4 \over 3}
&\Big\{& \str \Big[ 
F_{\m\n} D^\l D^\k F^{\n\r} D_\l F_{\r\s} D_\k F^{\s\m} 
+{1\over 4} F_{\m\n} D^\l D^\k F^{\m\n} D_\l F_{\r\s} D_\k F^{\r\s} \Big] \cr
&-& {2\over 15}\ ( f_{ace}f_{bde} +  f_{abe}f_{cde} ) \ \ 
\Big[ F^a_{\m\n} D^\l D^\k F^{b\n\r} D_\l F^c_{\r\s} D_\k F^{d\s\m} \cr
& & \phantom{{2\over 15} ( f_{ace}f_{bde} +  f_{abe}f_{cde} ) }
-{1\over 4}F^a_{\m\n} D^\l D^\k F^{b\m\n} D_\l F^c_{\r\s} D_\k F^{d\r\s} \Big] 
\ \ \Big\} \ .
\ea

\subsection{2 bosons / 2 fermions}

Finally we work out the mixed piece ${\cal L}_{\rm 2b/2f}^{\ap^4}$ 
much along the same lines as we did for the corresponding $\ap^3$ part.
We start with an ansatz similar to \eqn{bosfermap3} but with even 
two more derivatives to be distributed:
\ba\label{bosfermap4}
{\cal L}_{\rm 2b/2f}^{\ap^4}=
i g^2 \ap^4 & \Big\{&
\cb^a \g_\m D_\n D_\l D_\k \c^b\, \Big( y^{(1)}_{abcd}\ 
F^{c\m\r} D_\l D_\k F^{d\ \ \n}_{\ \r}
+ y^{(2)}_{abcd}\ 
D_\l D_\k F^{c\m\r} F^{d\ \ \n}_{\ \r} \cr
&&\phantom{\cb^a \g_\m D_\n D_\l D_\k \c^b\, }
+ y^{(3)}_{abcd}\ 
D_\l  F^{c\m\r} D_\k F^{d\ \ \n}_{\ \r} \Big)
\cr
&+&\cb^a \g_\m \g_\n \g_\r D_\s D_\l D_\k \c^b\, 
\Big(  z^{(1)}_{abcd}\ F^{c\m\n} D_\l D_\k F^{d\r\s}
+ z^{(2)}_{abcd}\ D_\l D_\k F^{c\m\n} F^{d\r\s} \cr
&&\phantom{\cb^a \g_\m \g_\n \g_\r D_\s D_\l D_\k \c^b\, }
+ z^{(3)}_{abcd}\ D_\l F^{c\m\n} D_\k F^{d\r\s} \Big) \Big\},
\ea
where again at the order $g^2$ we are working we can replace 
$D_\l D_\k \to D_{(\l} D_{\k)}$ etc, or equivalently replace $D\to \d$.
As for the order $\ap^3$ computation we can read the result of the 
amplitude computation from the order $\ap^2$ result \eqn{qftbosfermampl} 
by a series of substitutions which take into account the extra factors 
of $s$ and $t$ due to the additional derivatives. These are
\ba\label{substiap4}
y_{adbc} &\to&  {s^4\over 4} y^{(1)}_{adbc} +{t^4\over 4} y^{(2)}_{adbc} 
+{st\over 4} y^{(3)}_{adbc}\ , \quad
y_{adcb}\to {t^2\over 4} y^{(1)}_{adcb}+{s^2\over 4} y^{(2)}_{adcb}
+{st\over 4} y^{(3)}_{adcb} \cr
&&\cr
y_{dabc} &\to&  {t^4\over 4} y^{(1)}_{dabc} +{s^2\over 4} y^{(2)}_{dabc}  
+{st\over 4} y^{(3)}_{dabc}\ , \quad
y_{dacb}\to {s^2\over 4} y^{(1)}_{dacb} + {t^2\over 4} y^{(2)}_{dacb}
+{st\over 4} y^{(3)}_{dacb} \cr
&&\cr
z^+ &\to& {t^2\over 4} z^{(1)+} + {s^2\over 4}z^{(2)+} 
+ {st\over 4}z^{(3)+}\ , \quad
z^- \to {s^2\over 4} z^{(1)-} + {t^2\over 2}z^{(2)-}
+ {st\over 4}z^{(3)-}
\ea
where $z^{(i)\pm}$ are defined in analogy with \eqn{zplus} 
for $z^\pm$. We perform these substitutions in the resulting amplitude
\eqn{qftbosfermampl} and match the resulting expression to the string amplitude
$\sim uA+sB$. Vanishing of the $ \ub_1 \ks_3 u_4\  k\cdot\e\, k\cdot\e$ terms 
requires
\be\label{yoneytwoythree}
y^{(2)}_{abcd}=y^{(1)}_{abdc} \ , \quad y^{(3)}_{abcd}=y^{(3)}_{abdc} 
\ee
as well as six other conditions relating the $z^{(i)\pm}$ to 
combinations of the $y^{(i)}$, namely
\be\label{zyrel}
4 z^{(i)+}=-y^{(i)}_{adcb}-y^{(i)}_{dabc} \ , \quad 
4 z^{(i)-}=-y^{(i)}_{adbc}-y^{(i)}_{dabc} \ . 
\ee
This implies
\ba\label{calz}
{\tilde{\cal Z}} &\equiv& s^2 z^{(1)-} + t^2 z^{(2)-} + st z^{(3)-}
= t^2 z^{(1)+} + s^2 z^{(2)+} + st z^{(3)+} \cr
&=& -{s^2\over 4} \left(  y^{(1)}_{adbc} +  y^{(1)}_{dacb} \right)
-{t^2\over 4} \left(  y^{(1)}_{adcb} +  y^{(1)}_{dabc} \right)
-{st\over 4} \left( y^{(3)}_{adcb}+  y^{(3)}_{dabc} \right) \ .
\ea
Then the amplitude becomes
\be\label{fermbosampa4}
A_4^{\rm 2b/2f} \vert_{\ap^4} = -{i\over 4} g^2 \ap^4 {\tilde{\cal Z}}\ (uA+sB)  
= - 2i g^2 \ap^4 {\tilde{\cal Z}}\ K(u_1,\e_2,\e_3,u_4)\  .
\ee
Matching this to the string amplitude \eqn{A4ap4} implies
\ba\label{zmatch}
{\tilde{\cal Z}} &=& -{\pi^4\over 3} (s^2+t^2+st) \str \l_a\l_b\l_c\l_d \cr
&+& {\pi^4\over 90} \Big[ s^2(f_{abe}f_{cde}-f_{ade}f_{bce})
+t^2(f_{ace}f_{bde}+f_{ade}f_{bce}) + 2st(f_{abe}f_{cde}+f_{ace}f_{bde})\Big]\ .
\ea
Comparing \eqn{calz} and \eqn{zmatch} determines the combinations 
$ y^{(1)}_{adbc} +  y^{(1)}_{dacb}$ and $y^{(3)}_{adcb}+  y^{(3)}_{dabc}$
that are symmetric under simultaneous exchange of $a \leftrightarrow d$ and 
$b \leftrightarrow c$, much as was the situation at order $\ap^2$. There 
it was shown that the individual $y_{adbc}$ then are determined up to terms 
that lead to interactions that vanish on-shell and can be eliminated by 
field redefinitions. The same probably is true here, allowing to fix the 
ambiguities. Then the solutions to the matching equations are
\ba\label{yisol}
y^{(1)}_{abcd} &=& {2\pi^4\over 3} \str \l_a\l_b\l_c\l_d 
+{\pi^4\over 45} (f_{ace}f_{bde}+f_{abe}f_{cde}) \cr
y^{(2)}_{abcd} &=& {2\pi^4\over 3} \str \l_a\l_b\l_c\l_d 
+{\pi^4\over 45} (f_{ade}f_{bce}-f_{abe}f_{cde}) \cr
y^{(3)}_{abcd} &=& {2\pi^4\over 3} \str \l_a\l_b\l_c\l_d 
+{2\pi^4\over 45} (f_{ace}f_{bde}+f_{ade}f_{bce}) \ .
\ea
Finally \eqn{zyrel} is solved by
\be\label{zisol}
z^{(i)}_{abcd}=-{1\over 4} y^{(i)}_{abcd} \ .
\ee
Substituting these solutions back into our ansatz \eqn{bosfermap4} 
we obtain for the interaction
\ba\label{bosfermap4final}
{\cal L}_{\rm 2b/2f}^{\ap^4}
&=&
{2\over 3} i \pi^4 g^2 \ap^4 \str \Bigg\{
\cb \g_\m D_\n D_\l D_\k \c\,  
\Big( 
F^{\m\r} D_\l D_\k F^{\ \n}_{\r} + F^{\n\r} D_\l D_\k F^{\ \m}_{\r}
+ D_\l F^{\m\r} D_\k F^{\ \n}_{\r}
\Big) \cr
&&\phantom{{2i \pi^4 g^2 \ap^4\over 3}  }
-{1\over 4}\cb \g_\m \g_\n \g_\r D_\s D_\l D_\k \c\
\Big(
F^{\m\n} D_\l D_\k F^{\r\s} + F^{\r\s} D_\l D_\k F^{\m\n} 
+ D_\l F^{\m\n} D_\k F^{\r\s} 
\Big) \Bigg\} \cr
&+&
{i \over 45}\pi^4 g^2 \ap^4 (f_{ace}f_{bde}+f_{abe}f_{cde})
\Bigg\{
\cb^a \g_\m D_\n D_\l D_\k \c^b\ 
\Big(  F^{c\m\r}D_\l D_\k F^{d\ \ \n}_{\ \r}
+ F^{c\n\r}D_\l D_\k F^{d\ \ \m}_{\ \r} \Big) \cr
&&\phantom {{i \pi^4 \ap^4\over 45} (f_{ace}f_{bde}+f_{abe}f_{cde})}
-{1\over 4} \cb^a \g_\m \g_\n \g_\r D_\s D_\l D_\k \c^b\,
\Big( F^{c\m\n} D_\l D_\k F^{d\r\s} + F^{c\r\s} D_\l D_\k D^{d\m\n} \Big) 
\Bigg\}\cr
&+& 
{2 \over 45} i \pi^4 g^2 \ap^4 (f_{ace}f_{bde}+f_{ade}f_{bce})
\Bigg\{ \cb^a \g_\m D_\n D_\l D_\k \c^b\, 
D_\l  F^{c\m\r} D_\k F^{d\ \ \n}_{\ \r} \cr
&&\phantom{{2 \over 45} i \pi^4  \ap^4 (f_{ace}f_{bde}+f_{ade}f_{bce})}
-{1\over 4} \cb^a \g_\m \g_\n \g_\r D_\s D_\l D_\k \c^b\, 
D_\l F^{c\m\n} D_\k F^{d\r\s} \Bigg\} \ .
\ea

This completes our determinatioon of all terms in the open superstring 
effective action up to and including order $\ap^4 g^2$.

\section{Conclusions}

Higher-order in $\ap$ corrections to the low-energy effective action are 
of two types: additional field strengths $\ap g F$ or additional 
derivatives $\ap D^2$. An example of the first type is the famous 
$\ap^4 g^4 F^6$ term and an example for the second type are terms 
of the form $\ap^4 g^2 F (D^2 F) F (D^2 F)$ as obtained in this paper.
While it is usually argued that one can find
(interesting) situations where the former corrections are important, 
i.e. one has large fields, and the latter are small, i.e. slowly varying fields,
we have argued that both types of corrections are equally important. 

In the non-abelian case there is a formal argument which shows that the 
fluctuation spectra in such  backgrounds receive equally important contributions 
from both terms. We also presented a physical argument valid in the 
non-abelian and the abelian case. The basic idea is that large fields must 
fall off to zero at infinity. Either they fall off fast enough so that the 
fields are important only in a small region of space or they fall off 
slowly and are important over a large region. In the first case the 
derivatives are large and the higher-derivative terms are important. 
In the second case we showed that the total configuration necessarily 
has a large enough energy to form a black hole, so that gravity will 
couple in an important way to the Yang-Mills fields.

With this motivation in mind, we determined all corrections up to and 
including order $\ap^4$ as can be extracted from the open superstring 
four-point amplitudes. These terms all involve up to four Yang-Mills 
field strengths or fermions. They can be caracterised by being of 
order $g^2$ in the Yang-Mills coupling constant. There are the 
``four boson" terms involving four field strength tensors, the 
``two boson / two fermion" term involving two field strengths 
and two fermions and the four fermion term. In \cite{BBdRS}
all these terms  were determined at order $g^2 \ap^2$, and here we have obtained
all these terms in the effective action at order
$g^2 \ap^3$ (two extra derivatives) and order $g^2 \ap^4$ 
(four extra derivatives). They are given in eqs. \eqn{fullap3effaction},
\eqn{fourfermap4}, \eqn{fourbosap4} and \eqn{bosfermap4final}.

While at order $g^2 \ap^2$ all terms took the form of a symmetrised trace,
the order $g^2 \ap^3$ terms all are proportional to the product of two structure constants $f\, f$, so that they vanish in the abelian case. At order
$g^2 \ap^4$ and all higher orders, both, a symmetrised trace part and 
an $f\ f$ part are present. In particular, computations of the fluctuation 
spectra at order $\ap^4$ will have to take into account these explicitly 
non-symmetric $f\ f$-pieces of the higher-derivative order $g^2 \ap^4$ terms we 
have determined.


\vskip 1.cm 
\centerline{\bf Acknowledgements} 
 
I am grateful for fruitful discussions with my collaborators E. Bergshoeff, 
M. de Roo and A. Sevrin of the previous paper \cite{BBdRS}, as well as with
J.-P. Derendinger, R. Russo and K. Sfetsos. 
This work is supported by the Swiss National Science Foundation and by the
European Commission RTN programme  
HPRN-CT-2000-00131.


\appendix 
 
\section{Conventions and  identities} 

In this appendix we gather conventions and useful identities.
We use the same conventions as in \cite{BBdRS} but found it convenient
to collect them here again

\underline{Kinematics:} 
\be s=(k_1+k_2)^2 \quad , \quad t=(k_1+k_3)^2 \quad , \quad u=(k_1+k_4)^2 
\ee 
with all momenta incoming and we use signature $(+,-,\ldots ,-)$. Since  
all our states are massless we have $s+t+u=0$. 
 
\underline{Spinors:} 
The Clifford algebra is $\{\g^\m,\g^\n\}=2 \eta^{\m\n}$, i.e. $(\g^0)^2=+1$. 
Antisymmetric products of $\g$-matrices are defined with weight 1:  
$\g_{\m\n}=\ha (\g_\m\g_\n-\g_\n\g_\m)$ etc. Often used identities are 
\ba\label{gammaalg} 
\g_{\m\n\r} 
&=&\g_\m\g_\n\g_\r-\g_\m \eta_{\n\r}+\g_\n \eta_{\m\r} -\g_\r \eta_{\m\n}\cr 
\g_\m\g_{\n\r}&=&\g_{\m\n\r}+\g_\r \eta_{\m\n} - \g_\n \eta_{\m\r}\cr 
\g_{\n\r}\g_\m&=&\g_{\n\r\m}+\g_\n \eta_{\m\r} -\g_\r \eta_{\m\n} 
\ea 
The ten-dimensional spinors are 16-component Majorana-Weyl spinors and  
satisfy various identities. In particular, due to the Weyl property 
$\cb_1 \g_{\m_1\ldots \m_p}\c_2=0$ for all even $p$, and the expressions  
with $p>5$ are related to those with $10-p<5$. Due to the Majorana  
property anticommuting spinor fields satisfy 
\ba\label{majident} 
 \cb_1 \c_2 &=& \cb_2 \c_1 \ , \quad 
\cb_1 \g_\m \c_2 = - \cb_2 \g_\m \c_1 \ , \cr 
\cb_1 \g_{\m_1 \ldots \m_p} \c_2  
&=& (-)^p \cb_2 \g_{\m_p \ldots \m_1} \c_1 = 
(-)^{p(p+1)/2}\cb_2 \g_{\m_1 \ldots \m_p} \c_1  
\ea 
Note that when the anticommuting spinor fields are replaced by commuting  
spinor wave-functions we have the analogous identities but with an extra  
minus sign. 
 
There are also various Fierz identities which can be derived from the  
following basic identity \cite{BRW} valid for ten-dimensional  Majorana-Weyl 
spinors (a Weyl projector is implicitly assumed to multiply the r.h.s.) 
\be\label{genfierz} 
\p\lb = -{1\over 16} \g^\m ( \lb\g_\m\p) 
+ {1\over 96}\g^{\m\n\r} (\lb \g_{\m\n\r}\p)  
- {1\over 3840}\g^{\m\n\r\s\k} (\lb \g_{\m\n\r\s\k}\p)  
\ee 
from which follows 
\be\label{fierzone} 
\cb\g^\m\p \ \lb\g_\m\vf = \ha \cb\g^\m\vf \ \lb\g_\m\p  
-{1\over 24}  \cb\g^{\m\n\r}\vf \ \lb\g_{\m\n\r}\p  
\ee 
as well as 
\ba\label{fierztwo} 
\cb\g^{(\m}\p \ \lb\g^{\n)}\vf &=&  
-{1\over 8} \cb\g^{(\m}\vf \ \lb\g^{\n)}\p 
+{1\over 16}  \cb\g^{\r\s (\m}\vf \ \lb\g^{\n)}_{\phantom{\n)}\r\s}\p  
-{1\over 384}\cb\g^{\r\s\l\k (\m}\vf \ \lb\g^{\n)}_{\phantom{\n)}\r\s\l\k}\p\cr 
&& + \eta^{\m\n} \left[ 
{1\over 16} \cb\g^\r\vf \ \lb\g_\r\p 
-{1\over 96} \cb\g^{\r\s\l}\vf \ \lb\g_{\r\s\l}\p 
+{1\over 3840} \cb\g^{\r\s\l\k\tau}\vf \ \lb\g_{\r\s\l\k\tau}\p \right] \ , 
\ea 
where $(\m  \n)$ indicates symmetrisation in $\m$ and $\n$. 

\underline{Gauge group, $d_{abc}$ and $f_{abc}$ tensors :} 
We denote by $\l_a$ the hermitian generators of the fundamental  
representation of ${\rm U}(N)$. The various normalisations are fixed by  
\be\label{gaugegen} 
[\l_a,\l_b]=i f_{abc} \l_c \quad , \quad  
\{\l_a,\l_b\}= d_{abc} \l_c \quad , \quad  
\tr \l_a\l_b=\delta_{ab} 
\ee 
with real structure constants $f_{abc}$ and real $d_{abc}$. 
These definitions imply 
\be\label{threegentrace} 
\tr [\l_a,\l_b] \l_c = i f_{abc} \quad 
,\quad \tr \{\l_a,\l_b\}\l_c=d_{abc} \ . 
\ee 
The generators of the adjoint representation are  
$(T_a^{\rm adj})_{bc}=- i f_{abc}$, which is the only representation of  
interest to us. The covariant derivative then is 
\be\label{covder} 
(D_\m^{\rm adj})_{ac}= \delta_{ac}\d_\m - i g A_\m^b (T_b^{\rm adj})_{ac} 
=\delta_{ac}\d_\m + g f_{abc} A_\m^b 
\ee 
The field strength then is given by $[D_\m,D_\n]_{ac}= g f_{abc} F^b_{\m\n}$  
i.e. 
\be\label{fieldstr} 
F^a_{\m\n}=\d_\m A^a_\n - \d_\n A^a_\m + g f_{abc} A^b_\m A^c_\n 
\ee 
 
Possible 4-index tensors on the gauge group that could arise from a  
single trace 
are of the form $d_{abe}d_{cde}$, $f_{abe}f_{cde}$ or $d_{abe}f_{cde}$.  
There are 12 such possible tensors, but they are related by various 
Jacobi identities: 
\ba\label{jacobi} 
&&f_{abe}f_{cde}=d_{ace}d_{bde}-d_{ade}d_{bce} \cr 
&&d_{abe}f_{cde}+d_{bce}f_{ade}+d_{cae}f_{bde}=0 \ . 
\ea 
The first type of identities allows to express all $ff$ tensors as  
$dd$ tensors, and the second type of identities allows to express 3  
among the 6  $df$ tensors in terms of the 3 others. We may choose 
$\b_1=d_{abe}f_{cde}$, $ \b_2=d_{cde}f_{abe}$ and  
$\b_3=d_{ade}f_{bce}-d_{bde}f_{ace}$ as independent, and use them  
to express the three other $\b_4=d_{ace}f_{bde}$, $\b_5=d_{bce}f_{ade}$  
and $\b_6=d_{ade}f_{bce}+d_{bde} f_{ace}$: 
\be\label{dfidentities} 
\b_4=-(\b_1+\b_3)/2+\b_2\ ,\quad 
\b_5=-(\b_1-\b_3)/2-\b_2\ ,\quad 
\b_6=\b_1 \ . 
\ee 
Then, if we expand a general tensor as 
\be\label{gentensexp} 
X_{abcd}=x_1 d_{abe}d_{cde} + x_2 d_{ace}d_{bde} + x_3 d_{ade}d_{bce} 
+x_4 d_{abe}f_{cde}+x_5 d_{cde}f_{abe} + x_6(d_{ade}f_{bce}-d_{bde}f_{ace})\ , 
\ee 
knowing only $X_{(ab)cd}$ will leave $x_2-x_3$, $x_5$ and $x_6$  
undetermined, while knowing $X_{abcd}+X_{badc}$ will leave $x_4$  
and $x_5$ undetermined. 
Finally we note that 
\be\label{symtrdabc} 
\str\l_a\l_b\l_c\l_d= {1\over 12}  
\left(d_{abe}d_{cde} + d_{ace}d_{bde} + d_{ade}d_{bce} \right) \ . 
\ee 
 
\underline{Feynman rules:} From ${\cal L}_{\rm SYM} 
=\tr \left( -{1\over 4} F_{\m\n}F^{\m\n} +{i\over 2} \cb \g^\m D_\m \c\right)$  
we read the following Feynman rules for tree amplitudes (no ghosts):  
the fermion propagator is $+ i \delta_{ab}/ \ks$, the gluon propagator  
$-i \delta_{ab} \eta_{\m\n}/ k^2$ (any gauge dependent additional terms  
$\sim k_\m$ or $\sim k_\n$ drop out in all our amplitudes). All vertices  
are obtained from the relevant interaction terms with the rule  
$\d_\m \to - i k_\m$ where the momentum $k$ is going into the vertex.



\end{document}